\documentclass[journal]{IEEEtran}
\usepackage[utf8]{inputenc}
\usepackage{cite}
\usepackage[pdftex]{graphicx}
\usepackage{amsmath,amssymb,amsfonts}
\usepackage{algorithm}
\usepackage{algorithmic}
\usepackage{array}
\usepackage{amsthm}
\usepackage{textcomp}
\usepackage[english]{babel}
\usepackage{tabularx}
\usepackage{booktabs}
\usepackage{ragged2e}
\newcolumntype{R}{>{\RaggedRight\let\newline\\\arraybackslash\hspace{0pt}}X}
\usepackage{xcolor}
\usepackage{balance}

\title{Resilient Machine Learning for Networked Cyber Physical Systems: A Survey for Machine Learning Security to Securing Machine Learning for CPS\thanks{Authors are with the Data Science and Cybersecurity Center (DSC2), Department of Electrical Engineering and Computer Science, Howard University, Washington DC, 20059 USA. Corresponding Author  E-mail: danda.rawat@howard.edu. 
		
		This work was supported in part by the US NSF under grants CNS/SaTC 2039583, CNS 1650831 and 1828811, and by the DoD Center of Excellence in AI and Machine Learning (CoE-AIML) at Howard University under Contract Number W911NF-20-2-0277 with the U.S. Army Research Laboratory and by the U.S. Department of Homeland Security (DHS) under grant award number, 2017‐ST‐062‐000003.}
	\author{
		Felix O. Olowononi, Danda B. Rawat,\textit{ Senior Member, IEEE} and Chunmei Liu
	}}

\begin{document}

\maketitle
\begin{abstract}
Cyber Physical Systems (CPS) are characterized by their ability to integrate the physical and information or cyber worlds. Their deployment in critical infrastructure have demonstrated a potential to transform the world. However, harnessing this potential is limited by their critical nature and the far reaching effects of cyber attacks on human, infrastructure and the environment. An attraction for cyber concerns in CPS rises from the process of sending information from sensors to actuators over the wireless communication medium, thereby widening the attack surface.
Traditionally, CPS security has been investigated from the perspective of preventing intruders from gaining access to the system using cryptography and other access control techniques. Most research work have therefore focused on the detection of attacks in CPS. However, in a world of increasing adversaries, it is becoming more difficult to totally prevent CPS from adversarial attacks, hence the need to focus on making CPS resilient. Resilient CPS are designed to withstand disruptions and remain functional despite the operation of adversaries. One of the dominant methodologies explored for building resilient CPS is dependent on machine learning (ML) algorithms. However, rising from recent research in adversarial ML, we posit that ML algorithms for securing CPS must themselves be resilient. 
This paper is therefore aimed at comprehensively surveying the interactions between resilient CPS using ML and resilient ML when applied in CPS. The paper concludes with a number of research trends and promising future research directions. Furthermore, with this paper, readers can have a thorough understanding of recent advances on ML-based security and securing ML for CPS and countermeasures, as well as research trends in this active research area.
\end{abstract}


\section{Introduction}
The advent of the internet is a foundation for the birth of many of the developments and technologies that have significantly affected human, his interactions with others and the environment. The ability to electronically interconnect computer systems across the world made communication, collaboration and access to information easy and so serve as a tool for creativity and innovation. According to the International Telecommunication Union (ITU), broadband internet penetration is directly proportional to the employment and economic growth rates of a nation\cite{katz2012impact}. This is because the internet is the underlying technology for the digital revolution, which is responsible for paradigms and platforms like online commerce/shopping, online banking, online education, e-health and e-government. Beyond the internet, this digital revolution is also fueled by advances in wireless communications, proliferation of high capacity mobile devices, relatively lower cost of computing devices, alternative energy sources and access to larger memories. The field of parallel/distributed computing, cloud computing, quantum computing, nanotechnology and microelectronics and opto-electronics have also contributed immensely to these developments. In the early days of the internet, access was limited to computers and later on smart phones. However, further developments in wireless sensors have made it possible to incorporate minute and high capacity sensors into hardware devices used for everyday activities, thereby establishing connectivity to the internet. This development known as the Internet-of-Things (IoT) has made it possible to expand the internet from a worldwide network of computers to a worldwide network of computer and things, resulting in terms like Internet-of-battlefield things (IoBT), Internet-of-Vehicles (IoV) and Industrial Internet-of-Things (IIoT). IoT and wireless sensor networks (WSN) have pushed the frontiers of research in fields like manufacturing, transportation, healthcare, home automation, military warfare, entertainment and security \cite{jabbar2019design, mahmud2019smart, ramachandran2019internet, yang2016applications, jeschke2017industrial, abuzainab2017dynamic, azmoodeh2018robust}. \par
\textcolor{black}{CPS} leverage on the internet and WSN to act as intelligent systems that automate processes which were previously largely dependent on human efforts. Defined in so many ways by different authors, they fundamentally refer to physical and engineered systems where the monitoring, coordination, controlling and integration of the operations are done by a computing and communication core\cite{5523280}. CPS add a control action to the computing and networking dimensions of the IoT. As a result of the ability to use a feedback control to direct an actuator to take action based on physical measurements obtained from the sensors, the level of automation in CPS exceed that obtained from IoT systems. CPS have therefore generated a lot of interest from the industry, government and academia due to the immense potential they have to revolutionize virtually every field of human endeavor and solve practical challenges in our world.\par
One of the implications of the IoT and CPS when applied in critical infrastructure in agriculture, health, military, transportation, home automation and power systems is that a large volume of data is generated. This is because the devices are usually connected in real-time and remain continuously powered on. The concept of big data analytics (BDA) have therefore become applicable to IoT and CPS systems as they enable information to be accessible from data for making decisions such as fault prediction, diagnosis and predictive maintenance. In order to obtain value from the generated data, the use of data analytics to uncover hidden patterns, correlations and insights from large amounts of data are becoming increasingly popular as they introduce new functionalities to the systems highlighted above. Presently, CPS and other networked controlled systems are accountable for a large amount of data produced in the world. Various types of application that illustrate the concept of CPS are shown in Figure 1.

\begin{figure}
\includegraphics[width=\linewidth]{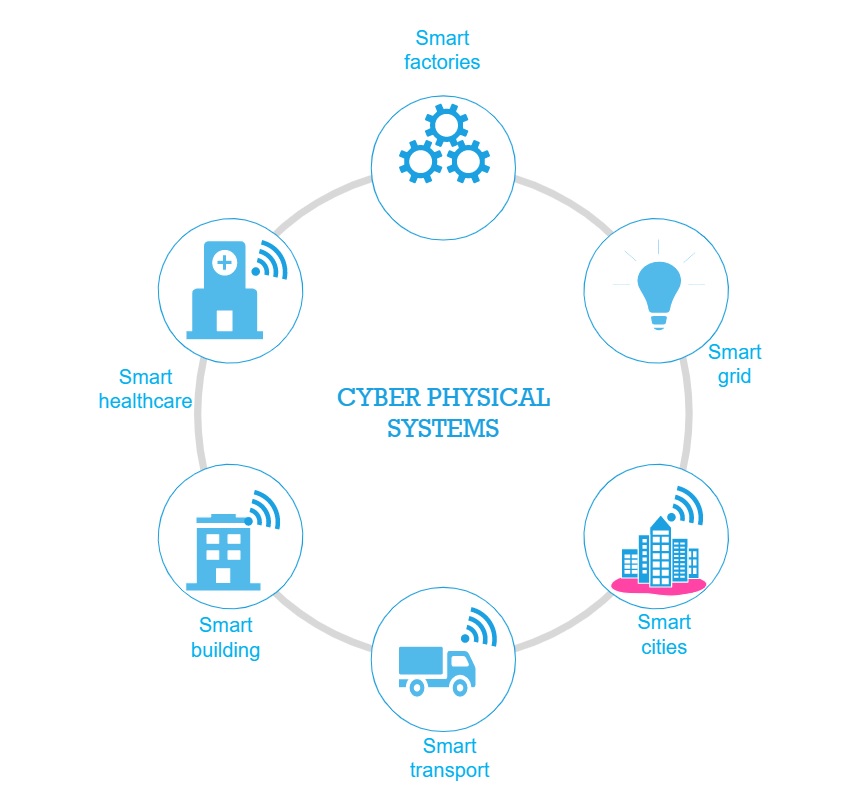}
\centering
\caption{Application scenarios of CPS}
\label{Figure 1}
\end{figure}

Despite all the hype of CPS, their actual deployment to solve real life challenges is hampered by safety and security concerns \cite{olufowobi2019saiducant}. CPS have very stringent requirements such as a need to operate in real-time and sensitivity to network challenges like delay. Moreover, the damage a failure causes to human life and infrastructure is more grave when compared to traditional information technology systems. Cybersecurity has become a dominant research topic in computer science and information technology. Initially limited to attacks on information using techniques like malware, adware, spyware and ransomware, security was guaranteed through the use of anti-viruses, firewalls and intrusion detection systems (IDS). The increase in inter-connectivity of sensors, actuators and controllers in CPS contribute largely to the rise of cyber attacks because they widen the attack surface and make these systems susceptible to adversarial activities. \textcolor{black}{Recently, ML and artificial intelligence (AI) algorithms have been used to enhance the efficiency of many systems\cite{li2019reinforcement, wickramasinghe2018generalization, dartmann2019big}. The reference of data and AI as ``new oil" and ``new electricity" respectively\cite{ng2016ai} underscore their impact in the present world.} Furthermore, emerging technologies like IoT, CPS, BDA and AI are the major technologies pushing for the fourth industrial revolution\cite{lasi2014industry, zhou2015intelligent}. As a result of these developments, a lot of research interest is directed towards the use of big data and data science principles to secure systems from \textcolor{black}{adversarial} attacks. The use of AI and ML for cybersecurity began with its implementation \textcolor{black}{in} IDS. Research in this area included malware and anomaly detection in information and communication systems. With the success recorded, ML was also used to achieve cybersecurity in IoT systems\cite{xiao2018iot, diro2018distributed, doshi2018machine, shakeel2018maintaining, 8302863}. Furthermore, the advent of \textcolor{black}{d}eep learning (DL) and \textcolor{black}{r}einforcement learning (RL) have contributed significantly to the deployment of ML algorithms to solve actual problems that posed a challenge to shallow algorithms and the more familiar supervised and unsupervised algorithms. Factors that support the use of DL in CPS include the high dimensional data generated and the continual growth of data\cite{wickramasinghe2018generalization}. Recently, researchers have combined DL and RL to arrive at the \textcolor{black}{d}eep \textcolor{black}{r}einforcement \textcolor{black}{l}earning (DRL)\textcolor{black}{; a development that has resulted in tremendous revolution in CPS research and continues to demonstrate a great potential to proffer solutions to present and imminent challenges}. This revolution is prominent in vehicular CPS like autonomous vehicles, because of the need to continually make decisions like lane changing and respond to traffic signs autonomously through image and pattern recognition\cite{ning2019deep, behzadan2018adversarial, 8324756, 8808950, 8500675}. \par 
There are however rising concerns with the use of AI in cybersecurity. Recent research has shown that systems that depend on ML algorithms for security are also prone to various forms of adversarial attacks. ML algorithms are data dependent and make inferences or predictions through data generated from various sensors in the networked systems. A predominant form of cyber attack in CPS and other systems therefore is the development of strategies to tamper with the data or the input. Consequently, the model is forced to produce wrong outputs. This is particularly common with neural networks, especially the \textcolor{black}{deep neural networks (DNN)} that have become very popularly used to secure CPS systems. Furthermore, the possibility of reusing the strategies meant to defend a system to attack it has also become a source of concern. The usage of AI and ML algorithms to defend systems can also be used by adversaries to attack the systems and perform adversarial attacks. Recently, it has been found \textcolor{black}{that} such attacks have a great potential because they are more sophisticated, faster and relatively cheaper since they leverage on the efforts of the defense systems to make themselves stronger and difficult to detect or curb. \par 
The possibility of compromising ML algorithms that are deployed to enhance cybersecurity in networked systems is a challenge that researchers must seek for ways to combat. In other words, it must be accepted that ML algorithms cannot totally prevent machines from gaining access to systems that are to be protected. As seen from \cite{barreno2006can, liao2011secure}, the concerns about the security of ML has been of interest for over a decade. Although many research have focused on IDS and spam e-mail filtering, not many have dwelt on what the systems must do in the presence of adversarial attacks. The desire to ensure that ML-enhanced systems continue to offer the services they should in the presence of adversarial attack led to the idea of resilient ML. 
This is a research domain that must be critically looked into to enhance the practical deployment of ML-based cybersecurity especially in CPS.

\begin{figure*}
	\includegraphics[width=\linewidth]{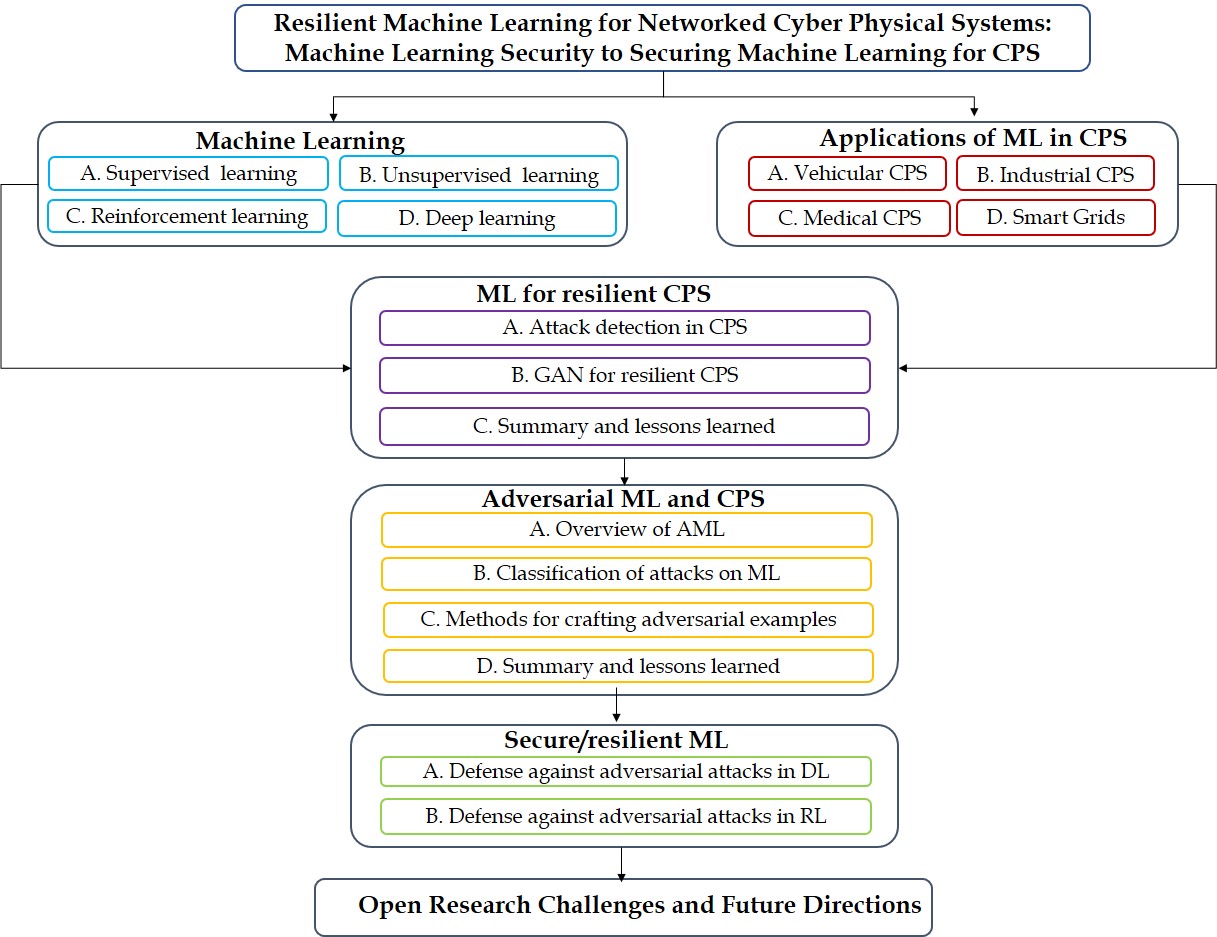}
	\centering
	\caption{Organization of the tutorial}
	\label{Figure 2}
\end{figure*}

\textcolor{black}{\subsection{Previous works}} 
In the last few years, a few survey papers have sought to discuss the issue of adversarial attacks and defense in ML models. Ozdag et al.\cite{ozdag2018adversarial} presented a sparse study on the subject. However, it was limited to DNN and the defenses were only focused on the Neural Information Processing System (NIPS) 2017 adversarial learning competition. Also, there was no focus on the resiliency of physical systems. A more comprehensive survey by Biggio et al.\cite{biggio2018wild} discussed the evolution of ML for over a decade. However, it was limited to \textcolor{black}{DL} and there was no attention on resiliency of CPS. Liang et al\cite{liang2019machine} focused on the good, bad and ugly use of ML for cybersecurity in CPS/IoT. The paper highlighted the advantages and disadvantages of ML applications in networked systems. However, it only presented a general overview of the subject. Finally, Yuan et al.\cite{yuan2019adversarial} presented an extensive survey on adversarial attacks and defense in DL. Just like the others, the authors did not dwell on RL, a commonly used approach in CPS. Furthermore, it was limited to methods published before 2017.
\textcolor{black}{\subsection{Contributions}}
This paper therefore aims to address the gaps of the previously highlighted research work by analyzing recent papers on adversarial attacks and defense, focus on both DNN and RL and also give insights on the subject with a focus on CPS. In summary, we comprehensively investigate the interactions between resilient CPS using ML and resilient ML when applied in critical systems. Specifically, the main contributions of this paper include:
\begin{itemize}
    \item Identify the roles of ML algorithms in the security and resiliency of CPS and prove from a comprehensive study of literature why the resiliency of ML algorithms must be a research concern.
  \item Present a comprehensive study of AML and \textcolor{black}{g}enerative \textcolor{black}{a}dversarial \textcolor{black}{n}etworks (GAN) in CPS and discuss their use both as an attack or defense against the resiliency of CPS.
   \item Present a discussion of recent trends, research challenges, insights and future research directions in the use of \textcolor{black}{ML} algorithms for achieving resiliency in \textcolor{black}{CPS}.
\end{itemize}

The remainder of this paper is organized as follows: Section II presents preliminary information on ML while Section III presents an overview of ML applications in CPS and specifically considers vehicular CPS, industrial CPS, medical CPS and smart grids. Section IV discusses the security and resiliency of CPS with ML while Section V focuses on AML in CPS. In Section VI, the focus is on secure and resilient ML. Section VII highlights open challenges and future research directions and section VIII concludes the study. \par 
\textcolor{black}{In an attempt to make the paper easy to read and navigate through, Figure 2 presents the structure of the survey.} 

\textcolor{black}{\section{Overview of Machine Learning}
In order to give the reader a good grasp of the discussion on the role of ML in CPS and the need to make ML models resilient to adversarial attacks, the various ML models commonly applied in CPS are briefly discussed in this section.}\par  
\textcolor{black}{One of the common definitions of ML is the ability of systems to ``make intelligent decisions without being explicitly programmed''. Despite the fact that it is used interchangeably with AI by some people, ML is actually a subset of the field of AI. ML approaches are data-driven in nature. The application of ML in CPS is therefore as a result of the large amount of data generated from the numerous sensors. Furthermore, ML techniques are usually categorized into three namely supervised, unsupervised and RL. Figure 3 shows the different types and the tasks that are carried out in each. These will be discussed in this section. Also, some of the algorithms that belong to each of these classes will be briefly discussed. This section will no doubt serve as a foundation to the major subject of this paper.}
\par

\textcolor{black}{\subsection{Supervised Learning}
In supervised learning, the training set contains the data samples and the desired solution or label(s). The goal of the ML algorithm is therefore to develop a function that maps the input to the output. After learning has taken place, an efficient model can take an unseen input and decide what the output should be. The most commonly used performance metrics for ML algorithms include accuracy, precision, recall and F1-score. The major task in this category include classification and regression. Supervised learning in CPS context are predominantly focused on classification tasks. A few of the most commonly used algorithms in CPS research are briefly discussed below.  }

\textcolor{black}{\subsubsection{Artificial Neural Networks (ANN)}
ANN are modeled after biological neurons in human brains. The perceptron is one of the simplest ANN architectures. Perceptrons can be trained to make predictions based on a fixed threshold. The multi-layer perceptron (MLP), obtained by combining many perceptrons together achieves better results. Activation functions like binary step, sigmoid and rectified linear unit ReLU) play the major role of converting an input signal of a node to an output signal. Technically speaking, they decide if a neuron should be triggered after mathematical computations. ANN models have been used to solve a lot of problems in our world. In CPS research, DNN's, which are discussed in a later section have become a choice model for solving various classification and regression tasks.}
\textcolor{black}{\subsubsection{Support Vector Machine (SVM)}
The capability of SVM to be used for classification, regression and even outlier detection tasks, generate accurate results and conserve computation power makes it an ML model of choice in CPS research. The algorithm has a goal of finding a hyperplane (decision boundaries) in an N-dimensional space that distinctly classifies the data points. The dimension of the hyperplane is dependent on the number of features in the dataset. For optimal results, the chosen plane should maximize the between data points of both classes. Prior to the widespread used of neural networks, SVM was a very popular ML algorithm for supervised learning.}	
\textcolor{black}{\subsubsection{k-Nearest Neighbors (kNN)}
KNN is a basic ML algorithm for classification (and regression) tasks. In CPS, it has been used for pattern recognition, data mining and intrusion detection. It's non-parametric behavior, which means that there is no need for assumptions on the data needed is a great attraction for its use in real-life applications. Basically, the algorithm decides the class of a test point based on majority voting by its K nearest neighbors. KNN makes predictions using the training dataset directly. For a new instance, predictions can be made by searching the entire dataset for the K most similar instances or neighbors and then summarizing the output variable for those K instances. Similarity between instances are measured using distance measure methods like Euclidean distance. Although kNN is no longer as popular as it used to be, some researchers still use it for their research.}

\begin{figure}
	\includegraphics[width=\linewidth]{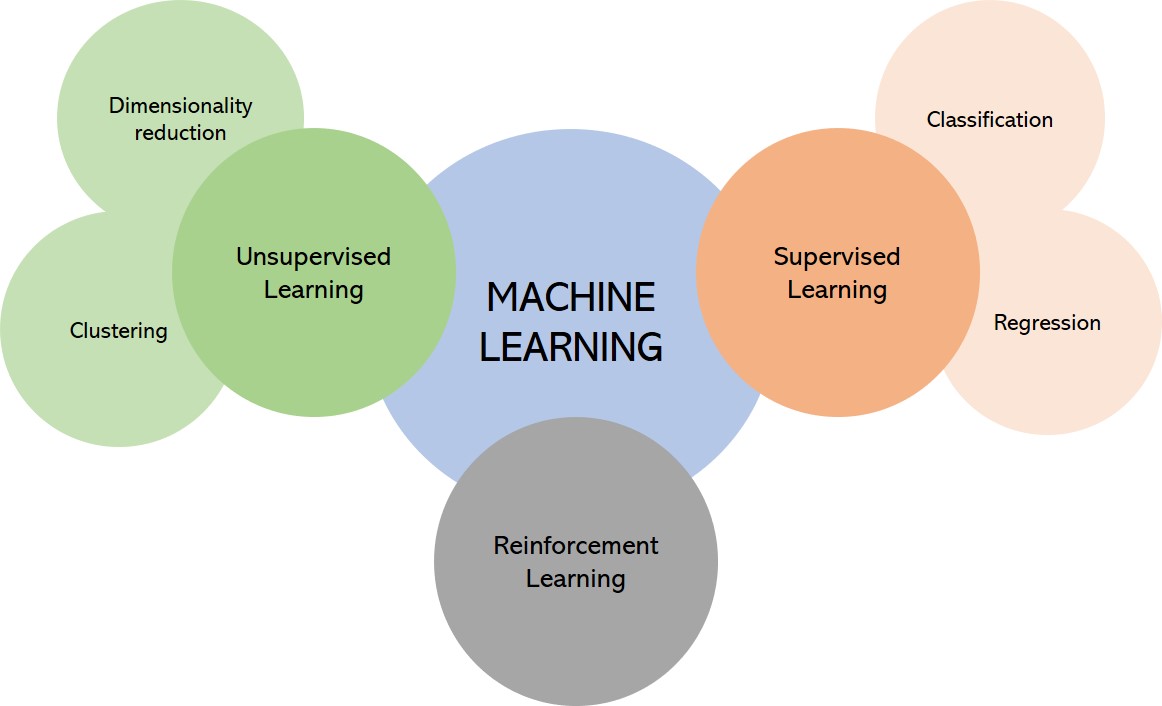}
	\centering
	\caption{Figure showing types of ML}
	\label{Figure 3}
\end{figure}

\textcolor{black}{\subsection{Unsupervised Learning}
In contrast to the supervised learning methods described above, unsupervised learning is achieved with unlabeled training data. The algorithm aims to learn or search for hidden insights in the unlabeled data. Tasks such as dimensionality reduction, clustering, density estimation, anomaly detection and visualization are achieved with this class of ML. Two of the most commonly used algorithms in this class, employed in CPS research are discussed below.}
\textcolor{black}{\subsubsection{K-Means Clustering}
K-means clustering is one of the simplest and popular unsupervised ML algorithms. Clustering refers to the task of identifying similar instances and assigning them to a group (cluster), after which underlying patterns can be identified. K-means clustering simply seeks to partition a number of datapoints into K clusters. Specifically, the algorithm operates by identifying K number of centroids (center of the cluster), assign every data point in the dataset to the nearest cluster with an ultimate aim to keep the centroids as small as possible. Although this method has advantages of being fast and scalable, it suffers limitations when the clusters have varying sizes and different densities. However, it has been used extensively in CPS applications for data analysis, dimensionality reduction, anomaly detection and even image segmentation.}	
\textcolor{black}{\subsubsection{Principality Component Analysis (PCA)}
PCA identifies the hyperplane that lies closest to the data and projects the data onto it. Put in another way, it is an orthogonal linear transformation method that transforms the data to a new coordinate system. The major advantage of PCA in ML research is its ability to guarantee efficiency in the ML lifecycle through a reduction in the number of features in the dataset, while still retaining the required information necessary for training. PCA, together with other dimensionality reduction algorithms like linear discriminant analysis (LDA) is therefore usually used to address the curse of dimensionality problem in ML.  }

\subsection{Reinforcement Learning (RL)}
In RL, the algorithm or agent learns to make decisions through its interactions with the environment. The algorithm performs this learning procedure in a trial and error manner, where it receives rewards and punishments for correct and incorrect performances respectively. The ultimate goal of the agent is to maximize the reward in any given situation.\par 
Figure 4 shows the interactions between the two major elements of a RL system; the agent and environment. While the agent depicts the algorithm itself, the environment represents the external condition or object the agent is acting on.  Four other important elements of a RL system include the policy, reward signal, value function and the model of the environment. The policy defines the behavior of the agent at any given time. This is usually achieved by mapping the states to the actions. The reward, which is the main goal of the setup is a function of the current action of the agent and the current state of its environment. The policy is usually changed by the agent to maximize the reward. The value function, though similar to the reward signal represents the long-term or cumulative  reward an agent can gather based on the states that are likely to follow the present state and the rewards associated to those future states. The model of the environment seeks to predict the behavior of the agent by making inferences about its next state and rewards based on information it has about a given state and action. \par 

\begin{figure}
	\includegraphics[width=\linewidth]{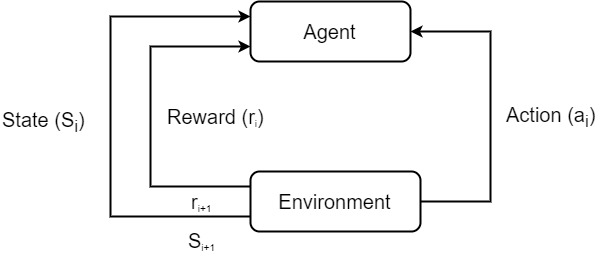}
	\centering
	\caption{Interactions between the agent and environment in a RL system}
	\label{Figure 4}
\end{figure}

A unique feature of the RL that distinguishes it from other types of learning algorithms is the trade-off between exploration and exploitation. This, in simple terms means that since the goal of the agent is to maximize the reward while interacting with the environment, it must seek to exploit the knowledge it already has based on past interactions and the rewards obtained. However, in order to select better actions in the future, the agent needs to explore other actions to maximize rewards. This occurrence is usually known as the exploration-exploitation dilemma.\par

Over the years, a number of RL algorithms have been developed. Q-learning, the classical algorithm was proposed by C.Watkins\cite{watkins1992q}. Next, the deep Q-Network (DQN)\cite{mnih2013playing, mnih2015human}, proposed by Google DeepMind in 2013 made popular the concept of DRL. Others include the value iterative network (VIN)\cite{tamar2016value}, asynchronous advantage actor-critic algorithm (A3C)\cite{mnih2016asynchronous}, trust region policy optimization (TRPO)\cite{schulman2015trust}, deep deterministic policy gradients (DDPG)\cite{lillicrap2015continuous}, proximal policy optimization (PPO)\cite{schulman2017proximal}
and the unsupervised reinforcement and auxiliary learning (UNREAL)\cite{jaderberg2016reinforcement}. Moreover, it is pertinent to state that with the introduction of the DQN, A3C and UNREAL, Google DeepMind has made a lot of impact on research in RL. In the coming sections, it will be evident that applications of RL in real life applications have leveraged mostly on the DQN. However, research into the defense of RL algorithms have attempted to study the DQN, TRPO and the A3C algorithms. Two of the most commonly applied RL algorithms in CPS are briefly discussed below.
\textcolor{black}{\subsubsection{Q-Learning}
This is the simplest and most commonly used RL algorithm. As the name implies, the goal of the RL agent is to learn the Q-Value, through iterative interactions with the environment and then use the information to take an appropriate action. The Q-Value, sometimes referred to as the Quality values is used to estimate the optimal state-action values. The Q-Value refers to the discounted accumulative rewards of an agent that starts with a state-action pair and follows a certain policy. At any state, the goal of the agent is to take an action with the largest Q-Value. Initially, the Q-Value is estimated to zero and then updated using the Q-Value iteration algorithm. However, a challenge of Q-learning that also affects its use in CPS is its inability to scale well to large Markov decision processes with many states and actions.}
\textcolor{black}{\subsubsection{Deep Q Network (DQN)}
To solve the aforementioned scaling challenge of Q-learning, DNN's are often used to estimate Q-Values. The DQN therefore introduced the subject of DRL. Put concisely, DQN is able to overcome the challenges of unstable learning with the following techniques; experience replay, target network, clipping of rewards and skipping frames. State transition samples generated through interactions with the environment are stored in a replay memory and consequently used to train the DQN. Furthermore, a target DQN is used to generate target values. The excellent results of the DQN algorithm have made them gain prominence in CPS research. This is especially evident in various tasks in autonomous vehicles.}	\par 
In summary, RL represents scenarios where there are interactions between an active decision-making agent and its environment, where the agent, without a knowledge of the environment however seeks to effectively achieve a goal in the environment. The agent relies on the fact that its actions can affect the future state of the environment and in essence the choices available to it in the future.
\textcolor{black}{\subsection{Deep Learning}		
 Recent research work has focused on the application of DL especially in the growing field of data science. The DL methods differ from the traditional shallow algorithms because they have several hidden layers, perform high level feature abstraction, generalize better on unseen samples and have shown to improve the performance of systems in which they have been deployed. These characteristics of DL have made them an attraction for various tasks in CPS. DL algorithms such as convolutional neural networks (CNN), recurrent neural networks (RNN) and autoencoders have been used for various tasks in CPS. }

\par

\section{ML Applications in CPS }
\textcolor{black}{CPS} are very important because of how they affect our daily lives. Their applications are found in critical infrastructure \textcolor{black}{as already highlighted in the last section}. Furthermore, ML models have been used to achieve various tasks in CPS. These include malware detection, intelligent resource allocation, detection of anomalous behavior, fault prediction, preventive maintenance and detection of attacks. \textcolor{black}{Figure 5 shows the contribution of ML algorithms to the advancement of the major CPS.} Since this paper is focused on security of CPS, more research on this will be presented in this section. The critical nature of these applications demand that they are safe and secure from attacks. However, this is not presently so. Some of the factors responsible for the challenge in securing CPS include the heterogeneous nature of the components, the complex interactions between the cyber and the physical sub-systems, and the widened attack surfaces generated by such interactions. Recent attacks on CPS have shown that the consequences are more pronounced, \textcolor{black}{especially when compared with} traditional information technology systems. However, any attempt to secure or defend CPS must begin with an understanding of the various vulnerabilities, threat and attacks that the system can suffer. \textcolor{black}{These are discussed for the major  CPS application scenarios presented in Figure 1.} In \cite{dibaji2019systems}, some prominent CPS cyber attacks were discussed. These include the Stuxnet attack, the RQ-170 attack, Ukraine attack, Maroochy Attack and the Jeep Hack. \par 
  In bolstering the argument for the use of DL in the security of CPS, the authors in \cite{wickramasinghe2018generalization} posited that they are able to handle the high dimensional data obtained from a large number of heterogeneous sensors in CPS. Furthermore, the ability to handle large data makes them improve because they are exposed to data with new vulnerabilities. In this section therefore, we introduce further the most common CPS applications and give an overview of the role of ML in these systems.
  
 \begin{figure}
 	\includegraphics[width=\linewidth]{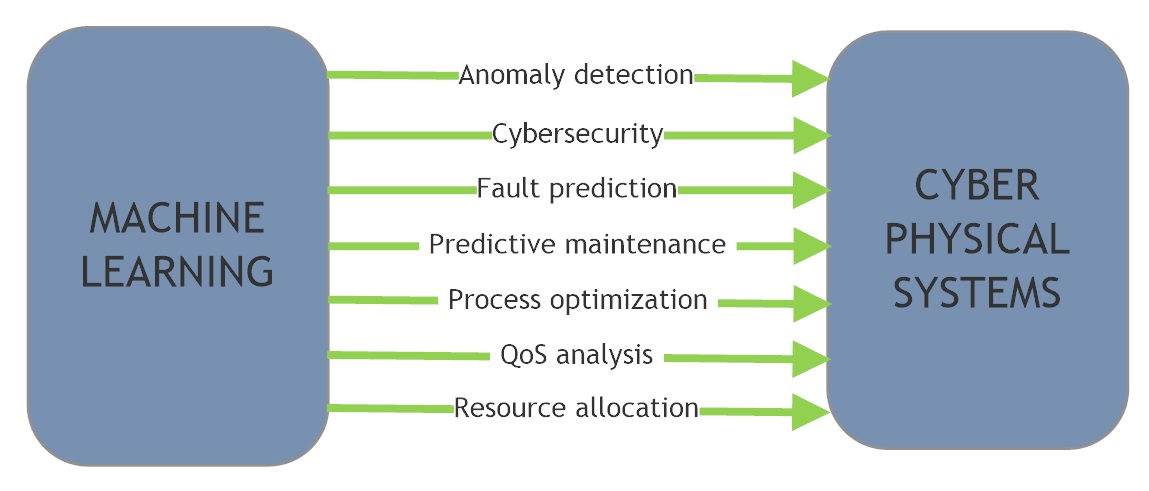}
 	\centering
 	\caption{Applications of ML in CPS}
 	\label{Figure 5}
 \end{figure}

\subsection{Vehicular Cyber Physical Systems (VCPS)}
The quest for a solution to persistent challenges like traffic congestion, vehicular accidents and its adverse effects on the environment has sustained the research interest in VCPS \cite{olowononi2020dependable}. Advanced driver assistance systems like cooperative adaptive cruise control and collision avoidance systems leverage on data from cameras, sensor networks and geographical positioning systems to increase the intelligence of vehicles. Vehicular adhoc networks (VANET) play a major role in the success of VCPS through the provision of interconnections between vehicles and road side units through wireless communication media. These are usually referred to as vehicle-to-vehicle (V2V) and vehicle-to-infrastructure (V2I) \textcolor{black}{communications}. Recent developments like self-driving cars and vehicular platooning are products of research advancements in vehicular technology. \textcolor{black}{Figure 6 illustrates how the components of a VANET communicate cooperatively to ensure the safe travel of autonomous and platooned vehicles. Since events that occur on the road are communicated to other vehicles, travel time and road congestion can be prevented because vehicles are able to take proactive decisions to ensure the comfort and safety of human and infrastructure.} 

\begin{figure}[b]
\includegraphics[width=\linewidth]{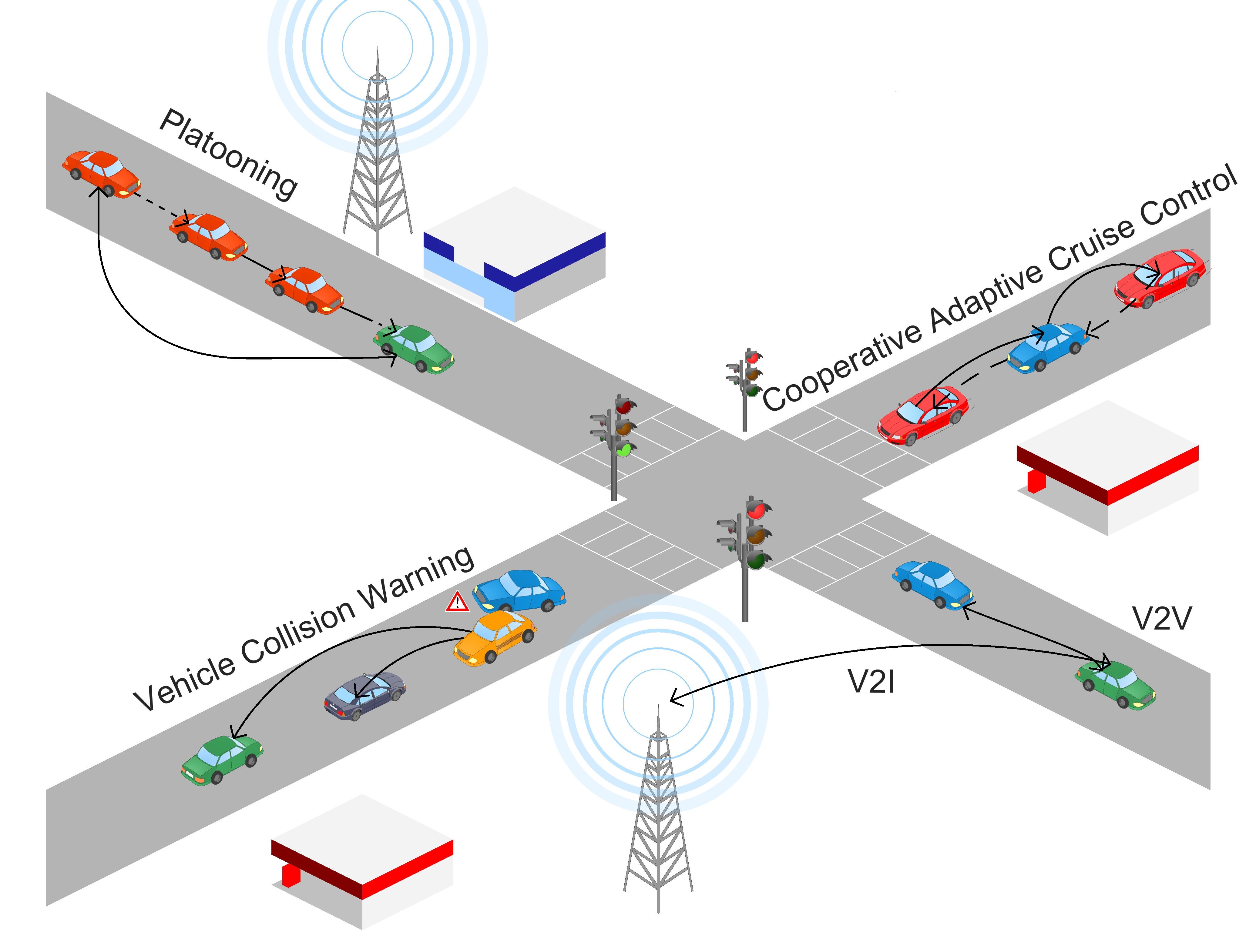}
\centering
\caption{Figure showing communications in a VANET}
\label{Figure 6}
\end{figure}

The security of VCPS have also been a major source of concern as they are subject to various adversarial attacks\cite{humayed2017cyber, giraldo2017security}. Initially, attacks on VCPS were studied from the communication perspective, with denial-of-service (DoS) \cite{8458117} and man-in-the-middle (MiTM) attacks accounting for the majority of attacks \cite{ahmad2018man}. Furthermore, \textcolor{black}{another kind of attack known as the replay attacks operates by delaying the messages sent in order to deceive the vehicles into taking action of belated information.} However, recently, attacks have also been launched on the control structure of VCPS. In this type of attack, the adversary alters a subset of control inputs, sensor measurements or control laws through replay and false data injections \cite{dadras2019resilient}. In the quest to defend VCPS from adversarial attacks, various schemes such as trust-based\cite{rawat2017securing}, blockchain \cite{8358773, 8753460, 8907054} and \textcolor{black}{ML dependent}\cite{olowononi2019security}  techniques have been used to secure vehicles from attacks. Recently, attention has been focused on autonomous vehicles where image and pattern recognition have been used to control the vehicles autonomously. Adversarial examples and \textcolor{black}{GAN} can be used to craft new inputs that cause the vehicle to take wrong decisions such as misjudging stop signs and other traffic signs.\par
This challenge of neural networks and its effect on CPS security is addressed comprehensively in a later section. Dominant research in securing VCPS with ML algorithms will continue to rise since the use of DNN's and other DL techniques are used to make decisions in vehicular systems more than other CPS. 

\subsection{Industrial Cyber Physical Systems (ICPS)}
CPS systems when applied in industries and manufacturing are usually referred to as ICPS, IIoT or Industry 4.0. In ICPS, a multitude of sensors, devices or agents deployed ubiquitously, and to remote locations in the plant connect over a communication network with other parts like actuators and controllers, for the purpose of monitoring, collecting, exchanging, analyzing and intelligently executing prompt actions on gathered data. Security and safety of ICPS is very important due to the nature of these systems and the cost and effects of adversarial attacks to human and infrastructure. A major case study of attacks to the ICPS is the Stuxnet attack on nuclear power plants in Iran. The novel manner the malware was distributed and the extent of damage that resulted from the attacks brings to the fore the issue of cybersecurity in network controlled systems. The malware was introduced into the network using a USB drive, after which the worm propagated itself to systems running on its target operating system. Moreover, it was also designed to search for the targeted component of the ICS; the high-speed centrifuges produced by Siemens. The primary goal of Stuxnet was to compromise the logic controllers of the system using  ``zero day" attacks. The developers then spied on the operations of the centrifuge for information, and launched attacks by taking control of the centrifuge. The unique qualities of Stuxnet were it's ability to evade detection systems during introduction and also remain hidden from human operators during the attack.\par
 ML has been used for fault prediction, identification of anomalous behaviors and predictive maintenance in industrial and manufacturing systems. \textcolor{black}{Haghighi et al\cite{haghighi2020machine} proposed an ML-based firewall for securing ICPS. The goal was to build on other researches that focused on accuracy to achieve zero false-positives in developed classifiers. The learning firewall receives labeled samples and performed self-configuration by writing adaptive preventive rules that avoid false alarms. Simulations on the KDD Cup'99 dataset showed that the proposed classifier could achieve zero false positives.}
 \textcolor{black}{Recently, edge computing have been used to improve the efficiency of ICPS. This is achieved by shifting tasks that are computationally intensive from edge devices with limited resources to high capacity edge servers. However, challenges of limited spectrum, low capacity batteries and lack of contextual information serve as a barrier to full realization of its potentials in industrial applications. ML approaches have been proposed to address these challenges. Sun et al.\cite{8863729} proposed a ML-enhanced method for offloading in edge devices in IIoT. The method is able to intelligently direct traffic to the edge server through the optimal path. Also, Liao et al.\cite{liao2019learning} proposed a learning-based context aware method for resource allocation in edge devices applied in IIoT.}\par
\textcolor{black}{From all of the research presented above, it is evident that ML algorithms will continue to play a great role for production efficiency in industrial systems. Moreover, in a bid to achieve resiliency in manufacturing plants, ML also have a great role to play through the incorporation of intelligence into such systems. Furthermore, as attackers begin to launch adversarial attacks in the ICPS, ML will also enhance the adaptive learning of the attack methods and hence proffer solutions that will mitigate such attacks. Specifically, the potential benefits of deploying various learning agents that will make the system self-configurable and resilient to adversarial attacks and operational faults will continue to make them of great research interest. However, recent studies has shown that ML algorithms will not be applied alone for achieving optimal results. They will be applied in hybrid with emerging technologies like software defined networks, blockchain and edge computing. Federated learning will also become suitably applied in ICPS due to the vast amount of sensors and heterogeneous devises in ICPS. Also, since industrial plants are very delicate, the increasing roles of ML algorithms necessitates adequate research into boosting the resilience of the algorithms themselves. This will no doubt, enhance the push towards the fourth industrial revolution.}

\subsection{Medical Cyber Physical Systems}
The field of medicine and healthcare has benefited a lot from the developments in information and communication technology \textcolor{black}{(ICT)} over the years. Some of the developments that have come from the impact of ICT on medicine and healthcare include advanced software enabled functionalities in medical equipment, continuous healthcare away from the hospitals to improve convenience both for the patient and healthcare givers, and increased connectivity of healthcare devices\cite{lee2010medical}. \par 
MCPS therefore refer to systems that can be used to remotely monitor vital signs such as heart rate, blood sugar and stress levels and automatically take actions to respond to situations when these vital signs are out of the normal threshold. This is usually achieved through the use of body sensor and wireless sensor networks. The most common examples of medical devices in MCPS include wearable devices and implantable medical devices (IMD) such as heart pacemakers and insulin pumps. Figure 7 shows the process of communication between the various units of a MCPS. \textcolor{black}{Sensors like electrocardiogram (ECG), electromyography (EMG) and blood pressure sensors implanted in the body of the patient measure signals from the heart, muscles and blood pressure respectively. These vital signs and readings are then sent wirelessly to the control unit. The sensors and control unit form a body area network (BAN), a wireless interconnection of computing devices through a wireless communication medium like redtooth. The control unit further relates the information to the medical server at the end of the medical personnel, through the access point and the internet. The medical personnel, on receiving this information is able to take proactive steps to prevent the occurrence of a heart attack even before it occurs.}\par  
However, from a security perspective, the wireless links between these medical devices, their controllers and the server make them susceptible to cyber attacks. The attacks could be either passive or active. In the passive attack, the hacker seeks to gain access to the data logged by the medical device, gain knowledge of the health conditions of the patient and use the information to blackmail or threaten such a person. However, in the active attack, the hacker's objective is to disrupt the operation of the MCPS and jeopardize the health of the patient. This is possible through jamming the wireless signals between the medical devices, thus resulting in DoS attacks that are dangerous to the health of the user. Moreover, an attacker might also compromise the sensors that measure the vital signals and cause it to give a wrong input to the controller. Consequently, when it responds to the false input, the controller directs the actuator to take an action such as pumping insulin into the blood stream, thereby putting the health of patient at risk. \par 
In summary, medical CPS due to the wireless interface over which they operate are susceptible to attacks like privacy invasion, jamming, noise, replay and false data injection attacks\cite{halperin2008pacemakers}. Furthermore, defects from software have also become a security concern for medical devices \cite{hanna2011take}. It is therefore evident that either through the hardware, software or the wireless communication through which they communicate, MCPS are vulnerable to various threats and thus efforts to secure them must be made a priority because of the threat they pose to human lives.

\begin{figure}
\includegraphics[width=\linewidth]{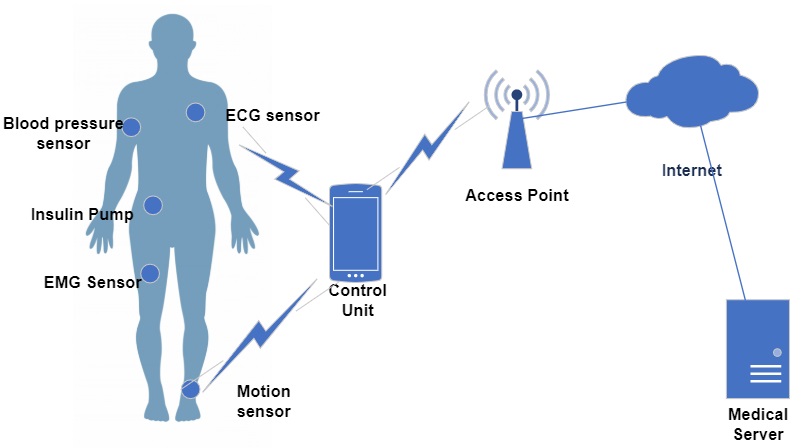}
\centering
\caption{Illustration of Medical Cyber Physical System}
\label{Figure 7}
\end{figure}

\subsection{Smart Grids}
The integration of the CPS concept to power systems resulted in smart grids. \textcolor{black}{According to the United States Department of Energy, a smart grid uses digital technology to improve reliability, security, and efficiency of the electrical system from large generation, through the delivery systems to electricity consumers and a growing number of distributed-generation and storage resources\cite{ton2015more}}. Smart grids make a lot of contribution to efficiency, and add a lot of functionalities to the generation, transmission and distribution of electricity. The smart grid comprises of two subsystems; the power application where the major functions of generation, transmission and distribution occur, and the supporting infrastructure where intelligent monitoring and control of these operations are carried out through the interactions of software, hardware and communication networks. \par 
Smart grid also suffer adversarial attacks when communications between field devices, control center and the smart meters are attacked through false data injection \textcolor{black}{(FDI)} and DoS attacks. \textcolor{black}{These attacks have largely been classified into attacks against confidentiality, integrity and availability.} Smart grids attacks usually result in blackouts that can cause a lot of damage to other systems that depend on it for power. The security of smart grids is therefore an active research domain. However, the complexity of smart grids and the heterogeneity of the CPS components have introduced significant difficulties to their security and privacy protection. The complex cyber-physical interactions pose a challenge to the assessment of threats and vulnerabilities in smart grids. Also, since attacks of power grids affect the efficiency of other dependent systems, hackers are not relenting in finding new loopholes to launch cyber attacks. \par
\textcolor{black}{Just like other CPS discussed, ML algorithms also contribute to the overall efficiency and security of smart grids. The data generated by the various sensors while the system is in operation is used to learn how to react to faults and attacks. To assert the invaluable roles of ML in smart grids, Zhang et al.\cite{zhang2018deep} presented a survey on the applications of DL, RL and DRL in smart grids. These include load forecast/power consumption\cite{8624311}, demand response\cite{pallonetto2019demand}, defect/fault detection\cite{lucas2020fault, bodda2019deep}, stability analysis\cite{moldovan2019detection} and cybersecurity\cite{hossain2019application}. From the security perspective, the authors in \cite{esmalifalak2014detecting} posited that recent attacks are stealthy and cannot be detected by traditional methods that depend on state estimation. They therefore proposed a ML-based approach for detection of FDI attacks in smart grids. The approach combines both supervised (SVM) and unsupervised learning (PCA). Khoda et al. identified that the ML algorithms used for securing CPS need to be resilient and proposed a novel adversarial retraining method for securing them. Without doubt, there is a great potential for ML to boost the resiliency of CPS to confidentiality, integrity and availability attacks. }

\begin{figure*}
\includegraphics[width=\linewidth]{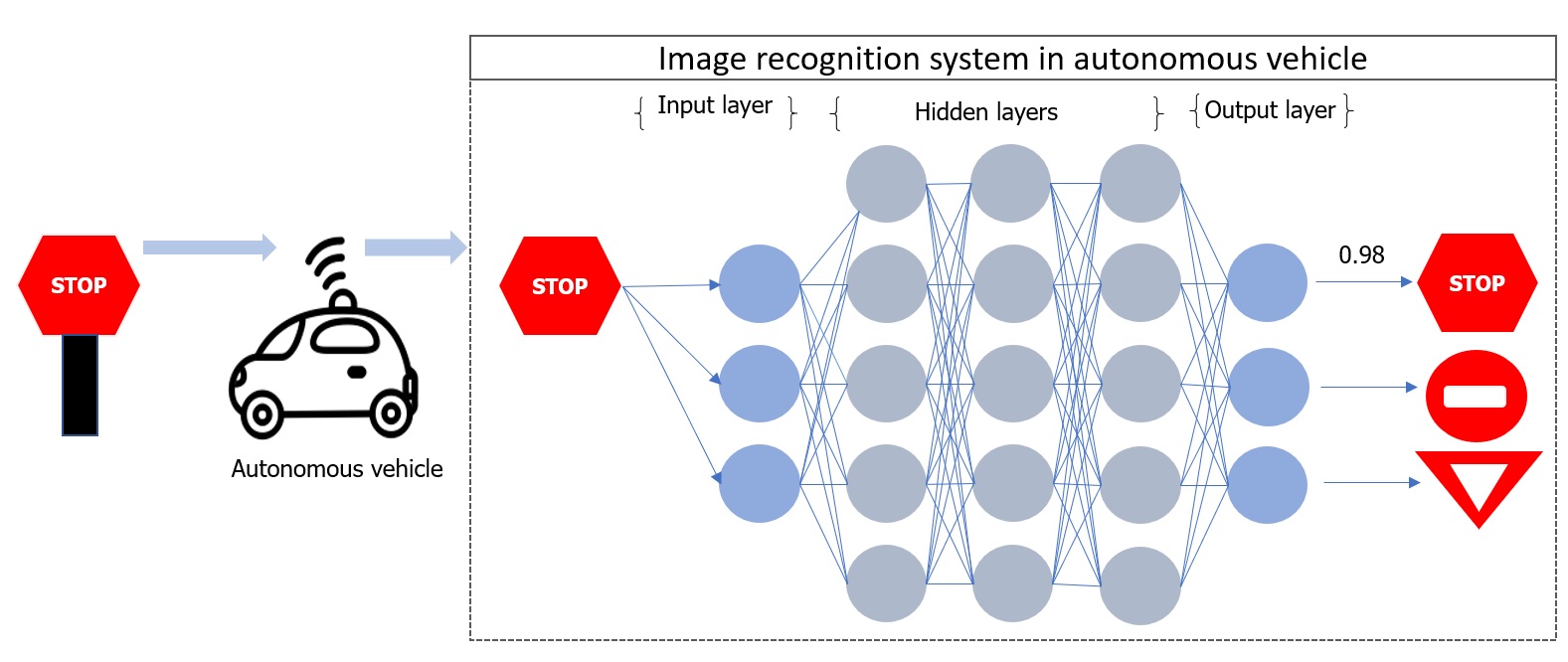}
\centering
\caption{Application of DNN in autonomous vehicles}
\label{Figure 8}
\end{figure*}

	\begin{table*}[]
	\setlength\extrarowheight{5pt}
	\centering
	\caption{Summary of research on Applications of Reinforcement Learning in Autonomous vehicles}
	\label{}
	\begin{tabularx}{\textwidth}{|R|R|R|R|}
		\toprule
		\textbf{Paper} & \textbf{Policy} & \textbf{Goal} & \textbf{Solution}\\
		\hline
		Automated speed and lane change decision making using deep reinforcement learning \cite{hoel2018automated}  & Deep Q-Network & Automated speed and Lane change decision& A DQN is trained to autonomously make decisions in self-driving vehicles. A CNN is also applied to high level inputs to quicken the learning process and optimize the agent’s performance.  \\
		\hline
		Learning Negotiating Behavior Between Cars in Intersections using
		Deep Q-Learning \cite{DBLP:journals/corr/abs-1810-10469} & Deep Q-learning & Intersection crossing &The vehicle observes distance and
		speed of vehicles on the intersecting road and use a policy
		that adapts its speed along its pre-defined trajectory to pass
		the crossing efficiently \\
		\hline
		A DRL Driving Policy
		for Autonomous Road Vehicles \cite{makantasis2019deep} &DDQN & Path planning for autonomous vehicle in a mixed driving environment (comprising of autonomous and manual driven vehicles) & The driving policy generates a collision-free trajectory for the autonomous vehicle to follow through dynamic programming. It operates by mapping the data from sensors on the AV and its environment to a goal. 
		
		\\
		\hline
		Trust-Evaluation-Based Intrusion Detection and Reinforcement Learning in Autonomous
		Driving\cite{8863727} & Q -Learning & Intrusion detection in autonomous vehicles based on trust evaluation & A Q-learning-based incentive mechanism to encourage autonomous vehicles report warnings to improve their trust values and utilities.\\
		\hline
		A DRL based trust management scheme for software defined vehicular networks\cite{zhang2018deep} & Dueling deep Q-network (DDQN) & Trust computation and path learning & The SDN controller is used as an agent to learn the most trusted routing path and determine the best routing policy.\\
		\hline
		Robust Deep Reinforcement Learning for Security and Safety in Autonomous Vehicle Systems \cite{ferdowsi2018robust}  & Q Learning & Robustness of AV to adversarial attacks& A novel DRL algorithm was proposed to maximize the robustness of AV dynamics control against data injection attacks. \\
		\hline
		Navigating Occluded Intersections with Autonomous Vehicles Using Deep Reinforcement Learning\cite{8461233} &Deep Q-Network&Intersection crossing in unsignaled intersections with occlusions& Improve safety of AV by analysing exploratory actions/creeping behaviours created by occlusions using DRL agents.\\
		\hline
		ML for Cooperative Driving in a Multi-Lane Highway Environment\cite{8734192} &Deep Q-Network&Investigate application of RL with cooperative driving in a highway environment& The use of information exchange in vehicular networking to control an AV in a multi-lane highway environment. \\
		\hline
		Scheduling the Operation of a Connected Vehicular Network Using Deep Reinforcement Learning \cite{8365853}  &Deep Q-Network&Improve safety and QoS in a connected vehicular network& Use DRL to train an agent to realize an energy-efficient and QoS-oriented scheduling policy. \\
		\bottomrule
	\end{tabularx}
\end{table*}

\subsection{Reinforcement Learning Applications in CPS}
The mode of operation of RL makes it very viable for improving the efficiency of CPS. In the last few, it has become a great tool for research in CPS. Significant impact of RL on CPS are highlighted in this section to reinforce this position. Kato et al.\cite{kato2018falsification} while stating that quality assurance in CPS remains a challenge as a result of factors like their heterogeneous and black-box components, proposed the use of RL to serve as a falsification approach in CPS. The goal of the trained RL agent is to learn the model behavior and then leverage on this information to compromise it for further investigations. Furthermore, the majority of the application of DRL have focused on image classification and recognition.\par
The application of RL in CPS have been aligned more towards \textcolor{black}{power systems} and intelligent transportation systems. \textcolor{black}{In power systems, the major research focus is on consumer cost optimization and other energy management endeavors. The advantage of RL lie in its ability to learn the best control policy and solve problems with a large state space. Kumar et al.\cite{kumar2019explainable} proposed the used of a DRL agent to operate in a variable pricing regime and learn to optimize the energy cost for the consumers. The agent manages the activities of the storage devices with a goal to maximize demand side cost savings. Other research in this direction are presented in \cite{yang2019large,8839066}. However, from the perspective of efficiency of the grid operations, Ren et al.  focused on the use of RL for load balancing in smart grids\cite{ren2019agent}. Highlighting the importance of developing cost-effective strategies for self-configuration and restoration of grid operations during blackouts, the authors identified a gap of other approaches to include the penchant to focus on maximizing the restoration efficiency and neglect the reliability-load balancing trade-off. They therefore proposed a method that uses the wolf pack algorithm (WPA), an RL strategy to optimize the reliability of the system during the restoration process. Moreover, in line with the context of this survey, Liu et al.\cite{liu2019reinforcement} differed from the other researches by proposing a method that leverages on RL for cybersecurity of power systems. Using a contingency analysis context, they leveraged on the Q-learning algorithm to develop an online learning scheme that models the activities of adversaries and the process of maximizing the attack strategy. The effect of the method was confirmed using simulations on eleven test cases.} \par
\textcolor{black}{In ITS, t}he advanced research in connected and autonomous vehicles have leveraged a lot on the field of ML and DL to increase the level of automation of vehicles and make them perform tasks that were previously performed by humans. In principle, sensors and other monitoring devices are now deployed in vehicles and other infrastructure to obtain data. The data gathered from  the sensors and devices is therefore analyzed for information that is used to make critical decisions on the road using DRL. Selected research on applications of RL in autonomous vehicles is presented in Table I. From a high level perspective, due to the level of uncertainty involved in autonomous driving of vehicles, DRL is used to carry out decisions such as intersection crossing, changing of lanes, speed control, trust computation and evaluation for safety and security. Furthermore, the Q-learning and deep Q-Network  are the most widely used RL policies applied in research in autonomous vehicles. \par 
Research into the application of DRL in autonomous vehicles and the other form of CPS is an interesting research area. It is expected that other decisions beyond the ones reported will be achieved with DRL. Furthermore, although the researches reported are theoretical, more needs to be done guarantee the safety and security of these systems if they will actually be implemented in real life scenarios. 

\textcolor{black}{\subsection{Summary and lessons learned}
	In this section, an overview of the roles of ML in CPS for four major application scenarios was discussed. The research trend showed that there has been a surge since 2017. We posit that the major factor responsible for this surge is the practical deployment of DL algorithms and their application in RL. Most research in this area have been directed towards VCPS and industrial applications. However, from a security perspective, it is also evident that the state-of-the-art methods for defending networked systems from attacks are no longer efficient due to the development of innovative attacks. Furthermore, the goal of most researchers have been to achieve optimal accuracy when simulations are carried out. However, issues of computational complexity and delay need to be brought to the fore due to the critical nature of CPS. The use of test beds will enhance the application of ML in CPS.   }

\section{ML for Resilient CPS}
In the last section, applications of ML in CPS was extensively discussed. In this section, in line with the focus of this paper, the goal will be to discuss the role of ML in cybersecurity of CPS.
Specifically, attack detection in CPS using ML algorithms and the role of generative adversarial networks in the resiliency CPS will be discussed.  
\subsection{Attack detection in CPS}
Attack detection in CPS is a dominant research topic because the early discovery of malicious behaviors or attacks will improve the chances of success of a counter attack to limit, mitigate or manage the extent of damage caused to the system. According to \cite{urbina2016survey}, attack detection schemes for CPS differ from traditional IDS for IT systems because of the additional physical dimension present in CPS. Security of CPS systems from malicious attacks have therefore been found to be more effective when the physics or physical properties of the systems are modeled and monitored. The authors in \cite{yan2019attack} suggested that performing attack detection at the physical domain of the CPS system serves as a last line of defense in the occasion that the other network layer schemes for attack detection are bypassed. The use of state space models like the Kalman filter, although commonly used in research for modeling dynamic systems, have been identified to suffer challenges like the inability to achieve optimal accuracy in complex CPS and the ineffectiveness in the detection of stealthy attacks\cite{miao2016coding}. The above listed challenges serves as an incentive towards the drive to applying ML schemes for attack detection in CPS. \par
In \cite{yan2019attack}, a ML-dependent attack detection scheme for CPS security was proposed. The success of the scheme began with a comprehensive feature generation scheme that leveraged on statistical, physical domain knowledge and DL techniques to generate features that better represent the non-linear and spatio-temporal relationships of the physical system. Furthermore, the combination of the generated features and the novel use of extreme learning machine for the detection model resulted in a high accuracy and also achieved early detection of malicious attacks in CPS. Furthermore, a behavior-based ML approach for detection in attacks in CPS was proposed in \cite{junejo2016behaviour}. Specifically, the authors focused on intrusion detection in the SWaT testbed. \textcolor{black}{In \cite{macasenhanced}, the importance of automatic detection of attacks and intelligent response in complex CPS was underscored. The authors highlighted that statistical process control methods for anomaly detection, such as cumulative sum (CUSUM) and exponentially weighted moving average (EWMA) are unable to produce effective results in networked CPS. This is due to their heterogeneous nature and the time series data generated from multiple sensors. They also posited that supervised ML techniques suffer from dearth of labeled data while unsupervised methods like clustering and temporal prediction methods have the challenge of  capturing temporal dependencies across different time series, coupled with the presence of noise in multivariate time series data from actual CPS operations. They therefore proposed an approach based on statistical correlation analysis between multivariate time series data and unsupervised DL algorithm for identification of adversarial operations in a complex multi-process CPS. Specifically, the approach uses a trained CNN autoencoder (CNN-AE) and convolutional long short term encoder-decoder (ConvLSTM-ED) models. The performance of the method was justified based on simulations carried out on the Swat testbed and comparison with state-of-the-art baseline methods.} Furthermore, Wang et al.\cite{wang2017detecting} developed a ML classifier for detecting time synchronization attack in CPS. Based on the principle of ``first aware", the results showed that the proposed classifier was able to detect direct and stealth time synchronization attacks. Shin et al.\cite{shin2017intelligent} proposed a DL-dependent method for detection of adversarial attacks in sensors deployed in autonomous vehicles. They also investigated the inertial measurement unit and wheel encoder sensors under conditions of uncertainty and non-linearity. \textcolor{black}{Also, the authors in \cite{ghafouri2018adversarial} used supervised regression as a means to detect anomalous sensor readings in CPS. By modeling the interaction between the CPS defender and attacker as a Stackelberg game, where the defender chooses detection thresholds in response to adversarial attacks, they proposed an algorithm for finding an approximately optimal threshold for the defender and proved that resilience can be boosted without sacrificing accuracy.}\par 
Due to the critical nature of CPS, it is required that they are dependable and secure. The dependability of a system entails availability, reliability, safety, integrity and maintainability. Also, security involves the common CIA triad; confidentiality, integrity and availability. Since CPS are ubiquitous, heterogeneous and complex in nature, it is possible for the operational conditions to change. The term ``resilience'' is therefore often used to describe the attributes of a system when it is resistant to malicious faults and persists in the delivery of its service or functions even when facing failure or adversarial circumstances. Also, it also refers to the persistence of dependability when a system is facing changes\cite{laprie2008dependability}. To underscore the importance of resiliency of next generation CPS, Barbeau et al. \cite{barbeau2019next} presented a vision for these systems. While acceding to the present case where increase in adversarial activities implies an increase in the likelihood of disruption of the system, they posited that by building mechanisms that leveraged on fuzzy decisions and ML, systems can continue to operate efficiently in such scenarios. However, in building resilient systems, the possible faults that pose as security threats must be considered. Furthermore, methodologies for detection and mitigating such threats must also be investigated.\par
The concept of adversarial networks became widespread when it was noticed that effective or competent adversaries are beginning to use certain strategies to evade detection systems which are designed to operate based on ML algorithms. Their activities which are generally classified as adversarial attacks are intended to attack the integrity, availability and privacy of the targeted system. In vehicular CPS, activities of an adversarial attack might include the flagging of a number of activities that are normal as an attack thus making the detection system unnecessary busy and affecting the availability of the system. However, the injection of false data into the system thereby causing it to make wrong classification is a dominant method for adversarial attacks in CPS.\par
As already discussed, resiliency in CPS will enhance their real-life deployment. Many systems have already been made resilient using AI and ML algorithms. Kannapan et al.\cite{kannappan2016incorporating} following in this direction proposed the incorporation of learning modules in ICPS as a viable method for achieving resilience. Agents monitor the activities of the system during normal operations and are able to recover to the learned states in the occurrence of a failure. 
In \cite{olowononi2019security}, a method for adversarial resiliency in VCPS was proposed. Furthermore, with regards to resiliency in CPS, Feng et al. sought to use the technique to ensure that a system maintains is activities in the presence of unknown cyber attacks. The authors first proposed a novel cyber state dynamics system that can dynamically and effectively ascertain the real-time impacts of present cyber attack and defense strategies. Next, they formulated the optimal defense problem as a two-player zero-sum game and finally developed a DRL algorithm to enable the scheme operate in real-time to suit its application in CPS. Simulations results reinforced their claim that the proposed DRL-based game theoretic actor-critic neural network was capable of learning the optimal defense and worst attack policies online accurately and in real-time\cite{feng2017deep}. Lokesh et al. in \cite{lokesh2015state} proposed a biologically inspired methodology for achieving resiliency in CPS using state awareness. Multi-agents deployed in the system are used to achieve state awareness. \par
In summary, research presented in this section show that ML algorithms have been successfully used both for attack detection and to buoy the resiliency of CPS.

\subsection{Generative Adversarial Networks (GAN) for resilient CPS}
\textcolor{black}{GAN's, as a result of their name have been widely portrayed as a technology that is solely used for compromising the integrity of various systems. However, in this section, we will show the contributions of GAN in the quest to improve the resiliency of CPS. This succeeds an introduction to the subject topic.} 
\subsubsection{Generative Adversarial Networks}
Generative models, which stem from DNN's typically operate by learning the density functions of original samples of data. They then use this information to craft fake samples that are not easily distinguishable from the real samples\cite{goodfellow2016deep}. They were described in \cite{yinka2019review} as an area of DL research that focused primarily on the generation of realistic data. The importance of generative models include the use of training and sampling models generated to ascertain the possibility of representing and manipulating high-dimensionality probability distributions, the potential of incorporating generative models into RL to improve decision making, their ability to both be trained with missing data and also make predictions even when their inputs have missing data, and the ability to enable ML work with multiple outputs\cite{goodfellow2016nips}. \par 
GAN's are a type of generative model that generate samples of a training dataset. In principle, they operate by setting up a game between the two major components; the generator and the discriminator. These two neural networks create samples that aim to possess the same distribution with the training data, and examine the veracity of the samples produced respectively. The goal of the generator is to deceive the discriminator to mis-classify while the discriminator learn through supervised learning. \textcolor{black}{Figure 9 illustrates the interactions between the two major components of a GAN.} The advantages of GAN's over other generative models include their ability to generate samples in parallel, relatively limited restrictions for the generator function, non-requirement of Markov chains and their ability to produce relatively better samples. The demerit of GAN's is therefore the need to find the Nash equilibrium of a game during training\cite{goodfellow2016nips}.\par

GAN's have become of interest in cybersecurity especially as DL continues to be a ML methodology of choice in recent years. Yinka-Banjo et al.\cite{yinka2019review} posited that the application of GAN's in cybersecurity is a developing research field. They authors in \cite{barbeau2019faking} also stated that beyond the possibility of GANs to generate fake data to fool a security system, they can also be used to defend systems. This is done by detecting the operation of adversaries through the generation and addition of fake samples to the training data to improve the robustness and resiliency of the system. The following section expand on this subject of discussion.

 \begin{figure}
	\includegraphics[width=\linewidth]{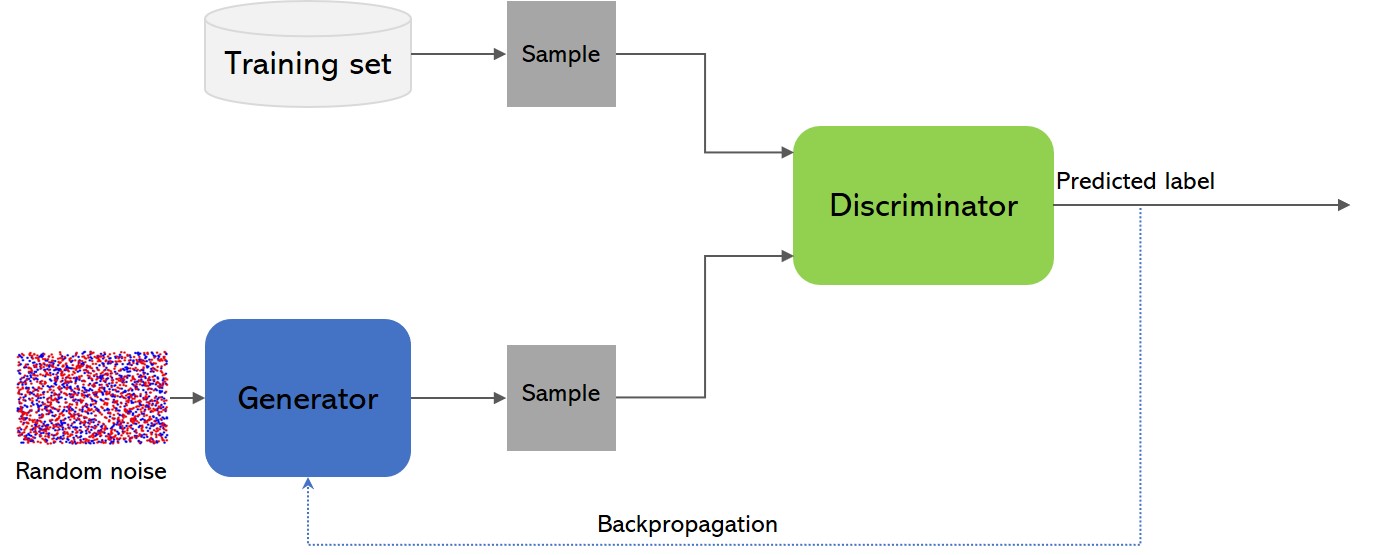}
	\centering
	\caption{Figure showing the operation of a GAN}
	\label{Figure 9}
\end{figure}

\subsubsection{The role of GAN's for resilient CPS}
GAN's have proven useful not only for attacks but for also defending CPS. The ability to use a generator and discriminator to craft adversarial samples can also be leveraged on to defend systems against attacks. CPS usually comprise of control loops where multiple sensors interact with the physical process or environment, micro-controllers that receive this data from the sensor networks and actuators that receive instructions from the micro-controllers in the form of electric signals and take a control action on the physical process. CPS therefore operate as a feedback control system. Although research initially focused on the communication aspect of CPS, the cybersecurity of the control and estimation processes in CPS have become a source of concern as adversarial attacks have also been targeted to disrupt them. To solve this challenge, a number of methodologies have been proposed by researchers\cite{7524933, li2017controllability}. However, ML models are now also being used to make the control process intelligent.\par The use of ML for learning and estimation in control systems have also made them susceptible to attacks suffered by systems that operate using ML. Adversarial samples can also be used to fool the system and thus compromise their integrity. To solve this problem, a novel conditional GAN (CGAN) was proposed in \cite{8714833}. The authors observe that the present controllers are unable to detect anomalous behaviors in the control loop, prevent such attacks or recover from an adversarial attack on the control loop. The system proposed therefore uses the conditional GAN to capture and learn the normal interaction of the physical system and the controller. The CGAN discriminator with the help of the discriminator captures the real behavior of the control loop during normal operation and thus able to identify anomalous behavior. The proposed system is also able to recover from the attacks by the prediction of the systems correct state. The results of the actual test proved that the proposed method is able to guarantee the detection and recovery from anomalous behaviors in vehicular CPS. Also, in \cite{li2018anomaly}, a GAN was used for modeling the distributions of the data streams of many sensors of a CPS operating in normal condition and another GAN to identify anomalies in the CPS, caused by attacks. The discriminator and generator therefore both use the multivariate time series data obtained form the sensors during normal operation to detect anomalies. The proposed model was validated using data from a Secure Water Treatment (SWaT). Other tasks such as predictive maintenance and fault diagnosis were highlighted as future areas where the propose algorithm will be applied to solve challenges in CPS in general. Furthermore, in \cite{li2019mad}, the combination on a GAN with a LSTM-RNN as the base model was used to detect anomalies in multivariate times series data generated from CPS. The authors, after testing their approach on the SWaT and Water Distribution (WADI) datasets  concluded that their method is effective in detective anomalies caused by cyber attacks in CPS. The use of GAN was also proposed for identifying security anomalies and cyber threats in the self-organizing networks of CPS\cite{belenko2018evaluation}. The authors as part of their future works intend to use the proposed model to secure a self-learning VANET/MANET. Chhetri et al.\cite{chhetri2019gan} proposed a conditional GAN security model that abstracts and estimates the relations between cyber and physical domains in ICPS and then analyses the security of the system. \par

From the research presented above, it is evident that ML and GAN, though predominantly seen as a cybersecurity concern to systems and networks can also be used to mitigate attacks to these systems. As research into the applications of AI in CPS continues to expand, GAN's will play a major role in the developments in future.
\subsection{Summary and lessons learned}
Without doubt, the role of ML in guaranteeing the resiliency of CPS is immense. The ability to train systems to take actions based on intelligent inferences from data will be used to enhance the efficiency and effectiveness of these systems. Ongoing research posit that ML algorithms will have a greater role to play in the quest to achieve real deployments of CPS.

\section{Adversarial Machine Learning (AML) and CPS}
Having already discussed the contributions of ML algorithms to the automation of various CPS, there are concerns that affect their successful deployment in CPS. AML involves the development of methods to compromise ML algorithms and their output, consequently influencing their ability to make right classification or predictions.  \textcolor{black}{Initially, most of the research on AML, especially for classification and pattern recognition tasks were generic in nature. However, since ML have been proven to be very instrumental in the progress of CPS, a number of researchers are beginning to explore the field of AML with focus on CPS. Rosenberg et al. \cite{rosenberg2020adversarial} in their study of adversarial learning in cybersecurity presented CPS and industrial control systems as a case study. Cai et al. \cite{cai2020detecting} studied an advanced emergency braking system for self-driving cars that operates by using DNN to estimate the proximity to an obstacle. They therefore used a regression model based on variational autoencoder to detect adversarial examples in learning-enabled CPS and concluded that the proposed method can detect adversarial examples effectively with a short delay. This was an improvement on their earlier study aimed at efficiently detecting out-of-distribution data capable of causing errors and compromising safety in CPS  \cite{ cai2020real}. The proposed method used variational autoencoders and deep support vector data description to learn models that efficiently identify and compute disparity between input data during the movement of a self-driving vehicle and the training set it was trained with. Similarly, the authors in\cite{ boursinos2020trusted, boursinos2020assurance} in a bid to complement the predictions of DNN in VCPS computed trusted confidence bounds for learning-enabled CPS. Furthermore, Li et al. \cite{ li2020searchfromfree} studied the challenge of adversarial attacks on ML models used for energy theft detection in smart grids. Their major contribution involved the development of a black-box attack that compromised meter measurements and consequently reports low power consumption measurements and so successfully fools the ML algorithm used for energy theft detection.  Clark et al. \cite{ clark2018malicious} also investigated the impact of adversarial attacks on the ML policies for controlling a robotic system. Finally, Xiong et al. in \cite{ xiong2020robustness} proposed attacks and defenses against ML algorithms used in learning-enabled controllers. } \par 
\textcolor{black}{The research presented above show that AML in CPS is an active research field and favorable results will facilitate the deployment of CPS in real-life scenarios.} In this section, a brief overview of the subject, classification of attacks peculiar to ML, methods for crafting adversarial examples in DNN and RL are discussed herein. 
\subsection{Overview of AML}
The field of AML brought to the fore the possibility of attacks on ML algorithms through the crafting of adversarial examples. The discovery that ML models (especially those based on neural networks like DNN's) can be fooled into mis-classifications with a high degree of confidence, by adding some perturbations to the training samples was first introduced by Szegedy et al. \cite{szegedy2013intriguing}. Furthermore, Papernot et al. suggested that the increased use of DL encouraged adversaries to deceive such systems where they are employed\cite{papernot2016limitations}. In a quest to explain this issue, Goodfellow et al. in \cite{goodfellow2014explaining} highlighted that speculations attributed the possibility of adversarial examples in DNN's to a combination of factors such as their extreme non-linearity, insufficient averaging of the model and insufficient regularization of the learning procedure. However, the authors disputed these speculations and posited that the linear behaviour of neural networks can be used to craft adversarial examples. On the basis of this discovery, they therefore developed a fast method for creating adversarial examples known as the fast gradient sign method (FGSM).\par
Over the years, interesting developments in the field of AML have followed since the discovery by Szegedy et al.  Beginning with various attempts to develop attacks which operate stealthily and have the greatest impact in as little time as possible, efforts are now geared towards developing defenses to these attacks.
\subsection{Classification of attacks on ML}
Attacks on ML models can be identified based on the goal of the attacker, stage of the attack and the level of information the adversary has on the targeted model. \textcolor{black}{Recently, the ability of an attack on a CPS to remain hidden has also become a source of research focus. Figure 10 therefore shows the various classifications of a attacks on ML algorithms.} 

 \begin{figure}
	\includegraphics[width=\linewidth]{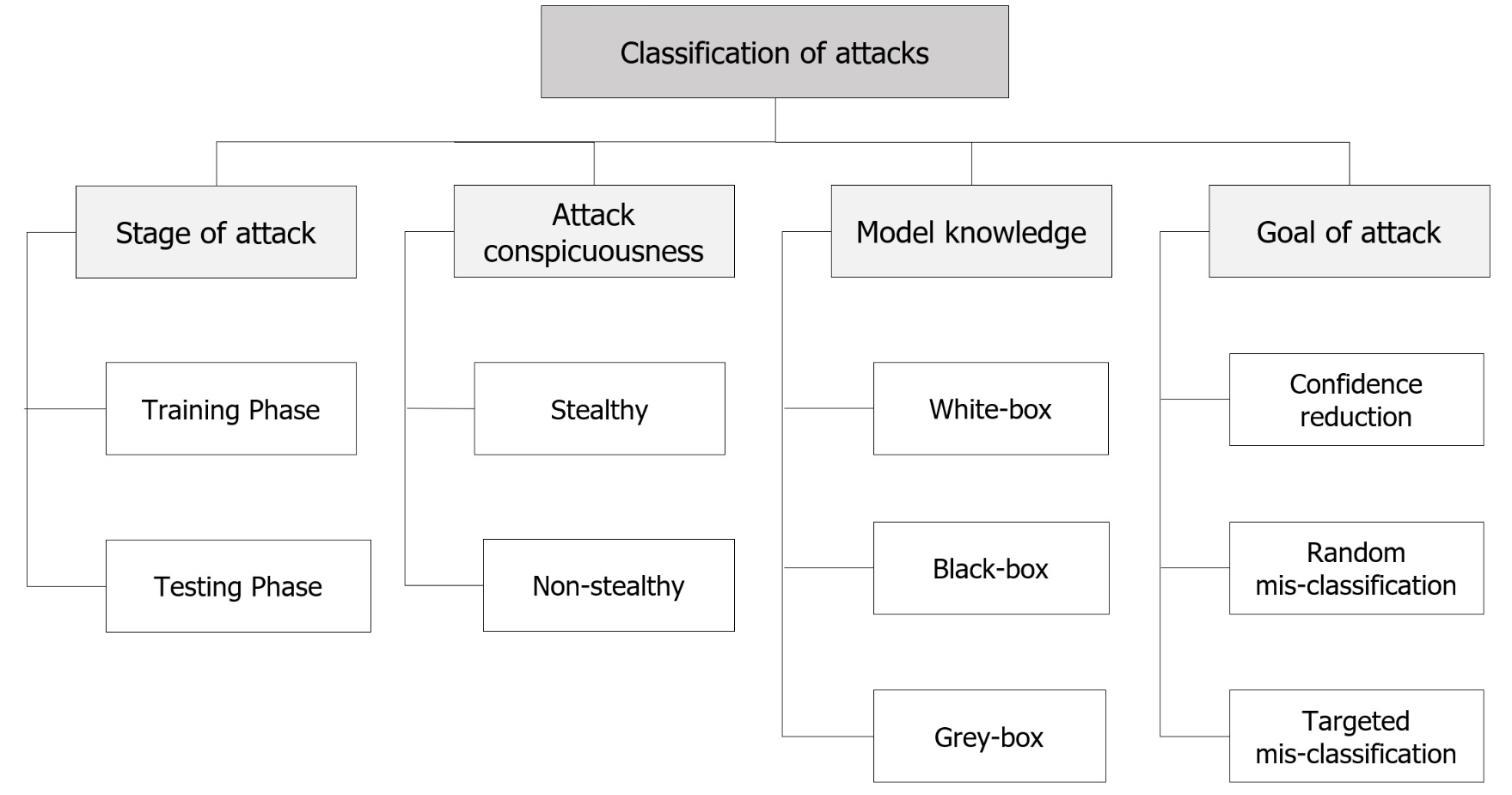}
	\centering
	\caption{Figure showing classification of attacks}
	\label{Figure 10}
\end{figure}

\subsubsection{Attacks based on the goal of the attacker}
A study of literature shows that there are different incentives for attacking a ML algorithm. In \cite{chakraborty2018adversarial}, three of these goals were given as confidence reduction, random mis-classification and targeted mis-classification. To reduce the confidence of the system, the attacker seeks to introduce an ambiguity in classification. However, the random classification occurs when the attacker changes the output classification to a random one different from the original. The targeted mis-classification occurs when the attacker seeks to supply the inputs or compel the system to produce and output class different form the original\cite{papernot2016limitations}. 
\subsubsection{Attack based on stage of the attack}
Attacks on ML models can occur either during the training or testing phases. However, as a result of its relative simplicity, majority of the attacks have been carried out during the training phase. The attack strategies employed in this method include the modification of data through FDI and logic manipulation\cite{liang2019machine}. Attacks during the testing stage however operate after the training has been completed. The goal of the adversary is to influence the model into making wrong classifications. \textcolor{black}{Figure 11 illustrates adversarial attack on a DNN in a ML-enabled CPS. Attacks during the training and testing phases are illustrated. Furthermore, inference attacks on the ML model is also shown.}

 \begin{figure}
	\includegraphics[width=\linewidth]{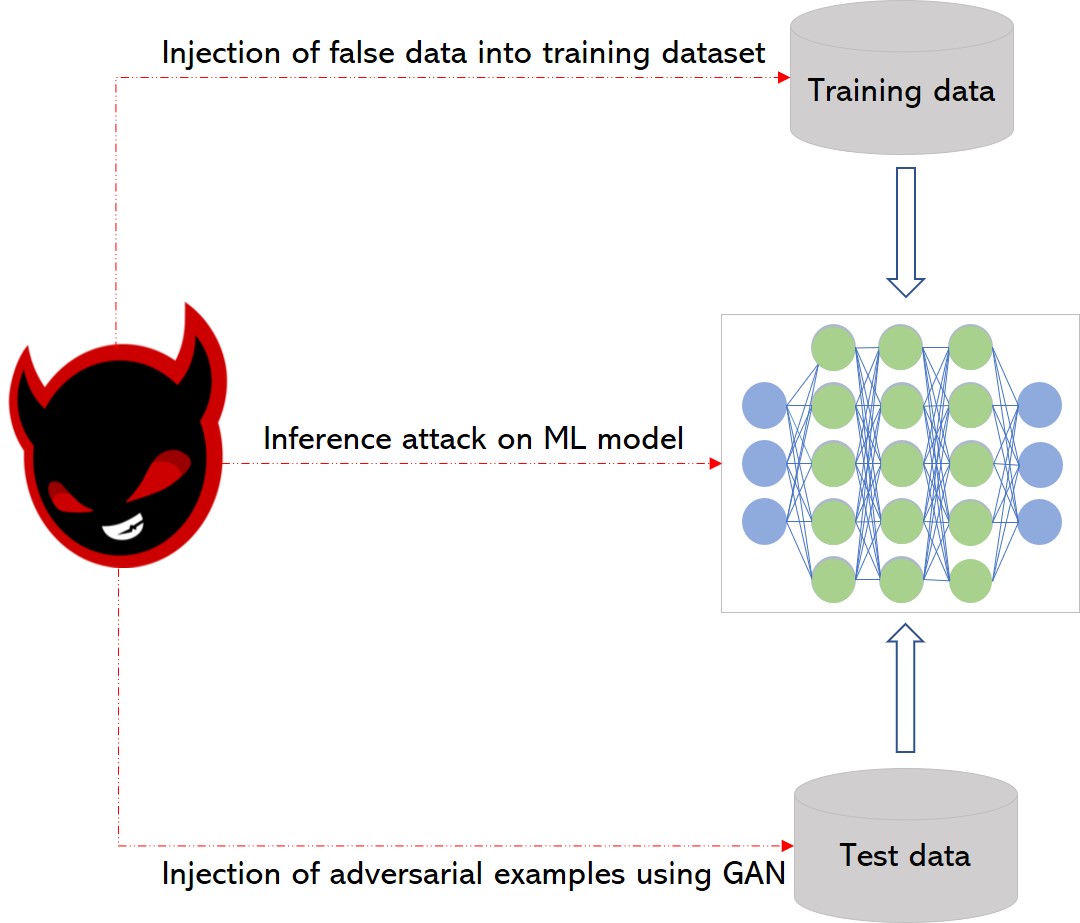}
	\centering
	\caption{Illustration of adversarial attack on ML-enabled CPS}
	\label{Figure 11}
\end{figure}

\subsubsection{Attack based on knowledge of model parameters}
Adversarial attacks have also been investigated from the perspective of the level of the knowledge of the internal architecture of the ML models. To this end, the primary classifications include white-box and black-box attacks. A hybrid of both of them, known as the grey box attack also exists. However, the focus here is on the white and black box attacks. 
\begin{itemize}
      \item White-box attacks: White box attacks represent a scenario where the attacker has a knowledge of the model parameters and the internal architecture of the model. This include information on the type of neural network, number of layers, and the number of neurons in these layers. Along with a knowledge of the learning algorithms and training process, the adversary seeks to modify these parameters. This therefore makes it a targeted adversarial method. Until recently, most studies in AML have focused on white-box attacks. However, it has been argued that in real-world systems, it is impractical to assume that the adversary will always have access to the information about the parameters of the target system because of the dynamic nature of such systems. There is therefore a high motivation for studying the attacks on ML models
      in scenarios where attackers do not have any information about the gradient function.
    \item Black-box attack: In this type of attack, the adversary has little or no knowledge of the internal structure or architecture of the ML model. The adversary therefore attacks the ML model by investigating the relationship between the input and output data sets. Consequently, the attacker either practices the attack action through the use of an agent learning model, or a direct manipulation of the input datasets that compels the model to perform mis-classifications. An approach for black-box attacks was proposed by Papernot et al. in \cite{papernot2017practical}. In this approach known as the transfer attack, the attacker begins by constructing a substitute model similar to the original model. Afterwards, the substitute model is attacked using the well known white-box attacks. The method proved not only to be functional in DNN but other supervised learning algorithms like logistic regression, SVM, decision trees and KNN. This approach was validated in \cite{liu2016delving, bhagoji2017exploring}. \par
    However, according to \cite{cheng2018query}, this approach of attacking the substitute model to perform a black-box attack has been faulted recently because studies has shown it usually leads to much larger distortion and low success rate. The authors in \cite{chen2017zoo} therefore proposed the score-based black-box setting that operates by querying the softmax layer. For further improvement of the query-based method, an autoencoder-based approach to reduce query counts and an adaptive random gradient estimation to balance query counts and distortion was introduced in \cite{tu2019autozoom}.\par
     Without doubt therefore, black-box attacks give a true picture of a real life scenario. Research into black-box attacks in adversarial learning will therefore continue to evolve in the coming years because since systems themselves are seeking to be resilient to such attacks, it will be very difficult for the attacker to have a knowledge of the internal architecture of the model.
    
\end{itemize}
\subsection{Methods for crafting adversarial examples in DNN}
In AML, the intention of the attacker is to generate a sample as similar as possible to the normal sample by adding the minimal perturbation to compromise the target model and also hide the change from human eyes. Such perturbations include fast gradient sign and other natural effects like fog and sunlight. Adversarial examples are a major tool used by an adversary to attack ML algorithms used in image classification.   According to \cite{ozdag2018adversarial}, studies in adversarial attacks are from three perspectives namely non-targeted adversarial attacks, targeted adversarial attacks and defenses against adversarial attacks. \par Generally, considering an input image $x$, the adversary seeks to find a minimum perturbation $\eta$ which when added to the the input image, produces an adversarial input $\overline{x} = x + \eta$ that can fool the system by causing it to mis-classify.\par

Studying the early works that brought to the fore the weakness and susceptibility of DNN's to adversarial attacks, it was seen that the various methods developed to detect the adversarial examples and make the ML model resilient to such attacks were dependent on the type of adversarial example and the method used to craft it. It was therefore common to see a method that proved to have a high accuracy in detecting adversarial examples fail in the future when tested with a newly developed adversarial example. From the foregoing, an understanding of the methods for crafting adversarial examples is therefore very pertinent to developing schemes for building resiliency against them. The state-of-the-art in the generation of adversarial examples have been classified into three namely one-step gradient-based approaches, iterative methods and optimization based methods\cite{dong2018boosting}. The gradient-based methods are the most popular of the three. They are of real interest in CPS security because autonomous vehicles and self-driving cars rely on image classification and pattern recognition to autonomously drive the vehicles and make important decisions as the vehicles travel on the road. A thorough understanding on how perturbations can be generated to make the systems take wrong decisions will enhance research in developing counter-measures to make them resilient to such attacks. Some of the well-known methods are briefly discussed in this section. Moreover, a summary of the peculiarities of the various approaches is presented in Table II.\par

	\begin{table*}[]
		\setlength\extrarowheight{5pt}
		\centering
		\caption{Summary of research on methods for crafting adversarial examples}
		\label{}
		\begin{tabularx}{\textwidth}{|R|R|R|R|}
			\toprule
			\textbf{Paper} & \textbf{Method} & \textbf{Contributions} & \textbf{Comments}\\
			\midrule
	Szegedy et al.\cite{szegedy2013intriguing}  & Solved penalized optimization problems & Introduced the concept of adversarial instability in neural networks; estimated adversarial examples by solving optimization problems &  The method was time consuming and did not scale well to large data sets\\
	\hline
	Goodfellow et al. \cite{goodfellow2014explaining} & Fast Gradient Sign Method (FGSM) & The method is fast and thus consumes less resources thus making the process of adversarial training a reality & The ease of generating adversarial examples have made it very popular. Initially used to test the effectiveness of most users but later found to be a relatively weaker form of adversarial attack.  \\
	\hline
	Kurakin et al.\cite{kurakin2016adversarial} & Basic Iterative Method (BIM) & Applies the FGSM method iteratively, with a reduced step size and clipping of pixel values after each step  &  It is an improvement on the FGSM, although the iteration sacrifices some level of speed\\
	\hline
	Papernot et al.\cite{papernot2016limitations} & Jacobian Saliency Map Approach (JSMA) & Also gradient based, the adversary, with a knowledge of the target model constructs adversarial saliency maps which it uses to detects input features that have the largest impact on classification of output and then attacks them with large perturbations & A targeted attack, the need to modify only a limited number of pixels in an input image makes it efficient \\
\hline
Carlini and Wagner\cite{carlini2017towards} & Carlini \& Wagner  & Three targeted gradient-based attacks; $CW_0$, $CW_1$ and $CW_2$ that were based on $L_0$, $L_2$ and $L_\infty$ norms respectively. More effective that the previous attacks & Robust against the defensive distillation method of adversarial defense and therefore highly recommended for testing methods for adversarial defense\\
\hline
Moosavi-Dezfooli et al.\cite{moosavi2016deepfool} & DeepFool & Improves on other methods by accurately computing the robustness of deep classifiers to adversarial perturbations especially in large data sets, and thus helps build more robust classifiers & This method computes a more optimal adversarial perturbation and used to show that adversarial training significantly increases robustness\\
	\hline
 Baluja et al.\cite{baluja2018learning} & Adversarial Transformation Network & A separate network is trained to attack the target network, and any input can be turned to an adversarial input. The advantages include fast and efficient training due to a single-forward pass, non-transferability and the choice of controlling the nature of mis-classification & A relatively new method that has not been well explored but has a great potential to be the new direction in this research area\\

			\bottomrule
		\end{tabularx}
	\end{table*}

\subsubsection{Fast Gradient Sign Method (FGSM)}
Goodfellow et al. posited that the susceptibility of DL models to perturbations is as a result of their linear behaviour. They therefore developed the FGSM for crafting adversarial examples. Introduced in \cite{goodfellow2014explaining}, adversarial perturbations to input images were crafted using the sign of the gradient or derivative of the models loss function with respect to the input feature vector. Furthermore, the authors in \cite{ozdag2018adversarial} described the FGSM as a method which creates an adversarial example by adding some weak noise to every step of optimization that approaches or moves away from the expected class.  The FGSM therefore seeks to fool the ML model into making wrong classification of the image through the addition of a small vector which is usually difficult to notice. \par
Just like the general case highlighted above, the minimal perturbation $\eta$ is obtained by perturbing each feature of the input image in the direction of the gradient. The mathematical expression is given below:
$$\eta = \epsilon * sign(\nabla_x J(\theta, x, y))$$
where $\eta$ represents the minimal perturbation, $\epsilon$ is a parameter that is used to determine the perturbation size, $J(\theta, x, y)$ the cost or loss function for training the DNN, $\theta$ represents the model parameters, $x$ is the model input and $y$ the targets to the model. The authors evaluated the performance of the method with MNIST and CIFAR-10 datasets. The results obtained confirmed that the method was able to fool DNN's to make wrong classifications.\par
Since the introduction of the FGSM, other methods to boost the generation of adversarial example using FGSM have been developed. This include iterative variants of the gradient based method\cite{kurakin2016adversarial}. Recently, it was argued that FGSM, being a one-step gradient-based method can only generate adversarial examples with high transferability when applied in a white-box model. A disadvantage is that it achieves a low success rate when trying to fool a black-box model. Since black box models are more practical in real world applications and systems now have their own defense mechanisms, it is important to develop attacks that can attack such models effectively without a prior knowledge of the internal architecture of the model. The solution proposed was the momentum iterative gradient-based methods to boost the examples,making it more transferable and achieve high success rates when applied in both white and black box models. The momentum based methods operate by iteratively accumulating a velocity vector in the gradient direction of the loss function with the principal aim of stabilizing update directions and shunning poor local maxima. This method is an improvement on the one-step gradient methods and the iterative methods. Readers who are more interested in this topic can refer to \cite{dong2018boosting} for a detailed understanding. Judging from the fact that the method won the first place positions in the targeted and non-targeted adversarial attack competitions in NIPS 2017, a knowledge of the method is very important.\par 
The FGSM since its introduction has become very popular because it is fast, simple and requires less computational resources thus making it very practicable. It has also been used to test the adversarial training method for enhancing resiliency of ML algorithms and helped to advance the research endeavors in this area. Adversarial training is discussed extensively in a later section. 
\subsubsection{Basic Iterative Method (BIM)}
Developed by Kurakin et al., it is an iterative variant of the FGSM\cite{kurakin2016adversarial}. The results obtained showed that the attacks was more effective than the FGSM attack. In this method, the adversarial noise $\eta$ is applied many times with a relatively small magnitude of the parameter $\epsilon$. One of the major benefits of this type of attack is the power it gives the adversary to control the attack. Furthermore, the BIM attack can be used by the adversary to successfully fool the network even when adversarial training is used to make the neural network robust. Adversarial training can increase the robustness of neural networks against a one-step FGSM attack but in a case where the attack is iterative, adversarial training will need to be adaptive to defend the model against the attack.. 
\subsubsection{Jacobian-based Saliency Map attack (JSMA)}
This type of attack also operates iteratively and focuses on targeted mis-classification. Proposed in \cite{papernot2016limitations}, the attack operates by using the forward derivative of a DNN to compel the model to classify into a predetermined class. The iterative nature of the attack makes it have a better success rate.

\subsubsection{Carlini and Wagner}
The Carlini-Wagner attack\cite{carlini2017towards, carlini2016defensive} attracted a lot of attention when it was introduced because of its ability to overcome the popular defensive distillation, which was proven to have the capability to overcome the FGSM attack. The approach has three attacks namely the $CW_2$ attack, $CW_0$ attack and $CW_\infty$ attack, based on the $L_2$, $L_0$ and $L_\infty$ norms respectively. These three attacks generated adversarial examples that successfully fooled neural networks using the defensive distillation into wrong classifications. The authors therefore recommended their approach as the relevant method for testing the effectiveness of any approach to be used in building models that are resilient to adversarial examples. 

\subsubsection{DeepFool}
The DeepFool method for crafting adversarial examples was proposed by Moosavi et al.\cite{moosavi2016deepfool}. DeepFool uses concepts from geometry to guide the search for the minimum perturbation needed to deceive a classifier to make wrong classifications. Furthermore, it uses the $L_2$ minimization method to search for adversarial examples. Through an iterative linearization of the classifier, the smallest perturbation needed to compromise the classification of samples is generated.

\subsubsection{Adversarial Transformation Networks (ATNs)}
Baluja and Fischer \cite{baluja2018learning} proposed a novel method for developing attacks for neural networks. The ATN operates by training a separate network to attacks another target network. In principle, adversarial examples can be generated by training the network to generate the perturbation to the input or an adversarial auto-encoding of the input. The possibility of generating both targeted and untargeted attacks and also executing training in a white-box or black-box manner makes the method attractive. Furthermore, ATN's have the advantage of granting the attacker the power to determine the nature of mis-classification that occurs in the target network and also reveal weaknesses in the target classifier.
It is pertinent to state that being one of the latest attacks developed, the ATN has not been subjected to enough discussion to prove its efficiency beyond the report of the authors. A comparative analysis of its ease of detection relative to other attacks will be instrumental in verifying its qualities. However, the qualities of training efficiency, need for single forward pass and its ability to convert any input into an adversarial example make it have a potential to contribute to research in AML.  

\subsection{Adversarial Attacks in RL}
The applications of DNN's in CPS, as previously studied are typically supervised as they majorly perform image classification and pattern recognition tasks. The security and resiliency of such ML algorithms have been discussed extensively in the previous section.
However, RL have recently become a very active research area. The ability of RL to achieve significant performance in various decision making tasks that involve uncertainty have endeared them to all. Results of endeavors by the industry and academia show that it has solved a myriad of problems, especially in VCPS research. The attacks on RL algorithms differ from those on DNN's. However, there is some intersections because of the use of DRL. Specifically, DNN's are used to approximate the action-value function. The policy, because of its role in directing the agent on the most effective or rewarding action to take in response to the state of the environment is usually the target of many attacks. Moreover, it is pertinent to state that although not much attention has been given to adversarial attacks in DRL in the past, this is bound to change in the coming years as DRL continues to extend from the initial video games to more critical systems like robotics and autonomous vehicles. \par 
The state-of-the-art approach in research into security and resiliency of ML models usually begins with the development of attack and threat models to ascertain their ability to compromise the model.
\textcolor{black}{Adversarial attacks on DRL can be classified into three. These include those that target the reward by perturbing it directly or the reward signals through the states; attacks that  target the DRL policy by perturbing the states, perturbing the environment, involving an adversarial agent and model extraction attacks; attacks that target the observation by perturbing the states; and attacks that target the environment \cite{ilahi2020challenges}.}
One of the earliest study in adversarial attacks in RL was carried out by Behzadan et al. in\cite{behzadan2017vulnerability}. The research affirmed that the policy and induction in DQN, a common type of RL technique are vulnerable when adversarial examples are introduced into the input, and that transferability of adversarial examples is also possible from a DQN model to another. The attack mechanism developed by the authors depended on previously discussed methods for crafting adversarial examples like FGSM and JSMA. The policy induction attack presented has an adversary that trains a DQN with an adversarial reward, and then
uses the trained policy to craft targeted adversarial examples that attract the agent to take actions leading into obtaining the adversarial reward.\par 
Furthermore, Huang et al. \cite{huang2017adversarial} also towed the same line to show that existing adversarial examples crafting techniques can be used to limit the performance of RL policies. The authors state that although the research presented in \cite{behzadan2017vulnerability} focused on adversarial attack at the training phase of the agent to prevent learning, their work was an improvement because it presented adversarial examples at test time to investigate adversarial attacks on an RL agent. It is pertinent to also state that the study was based on white-box and black-box settings. The authors in \cite{kos2017delving} also used the FGSM attack in their study on adversarial attacks in DRL policies. The major contributions of the research include a comparative analysis of adversarial examples and random noises in attacking DRL policies, and the exploration of the value function of the policy as guide in the injection of perturbations. This novel method has the advantage of reducing the time the adversary expends in injecting examples to record a successful attack and therefore increases the probability of it remaining undetected. Lin et al. \cite{lin2017tactics} also proposed the limitation of the time of operation of an adversary as a method to guarantee the stealthiness and efficiency of adversarial attacks in DRL agents. To this end, using the C\&W method \cite{carlini2017towards} for crafting examples, they proposed the strategically-timed attack and the enchanting attack. The strategically-timed attack operates by determining the most effective time an adversarial example should be crafted and selectively attacks at a subset of the time steps as against the usual uniform attack presented in \cite{huang2017adversarial}. The enchanting attacks combines a generative model for the prediction of future states and a planning algorithm for the generation of predetermined actions to lure the agent to a desired target state after a number of steps. Significant among the results presented is that the strategically-timed attack can achieve the same effect as the uniform step attack, despite the agent being attacked four times less. \par 
Sun et al.\cite{sunstealthy} also predicted the critical point attack and antagonist attacks. Just like the attacks in \cite{lin2017tactics}, both attacks operate by building a model to predict the future states of the environment and the action of the agent and then select that in which maximum damage can be achieved in minimum steps. The C\&W approach for generating adversarial attacks is also used in the research. The authors in \cite{hussenot2019targeted},while critiquing the works in \cite{huang2017adversarial, kos2017delving} as unrealistic for ignoring the dynamic nature of RL models by assuming that attacks can be generated per state proposed the targeted attacks on DRL agents through observations. They also argued that the general perception in the adversarial RL community that the goal of attacks is to largely achieve a drop in the performance of the model is untrue  as it may also be to lure the agent into an action determined by another policy. They therefore proposed the per-observation attack and the universal-masks attacks using the FGSM method of crafting adversarial examples. \par 
Deviating from the use of the FGSM and C\&W methods for crafting adversarial examples, Tretschk et al.\cite{tretschk2018sequential} in studying adversarial attacks in RL were the first to use the ATN to learn to generate the attack against the policy network. Their goal was to show that unlike the state-of-the-art, a sequence of attacks can be used to thrust a random adversarial reward on the policy of the target system. The effect of this is that the target agent can be deceived to optimise for the adversarial agent as a result of the sequential attacks. This approach, though similar to those in \cite{behzadan2017vulnerability, lin2017tactics} is unique due to the use of the ATN and the application of the attack at test time. \par
With a goal to investigate the application of DRL in CPS, Lee et al.\cite{yeow2019spatiotemporally} proposed the white-box Myopic Action Space (MAS) and the white-box Look-Ahead Action Space (LAS) attacks. In contrast to other works which have focused more on attacking the RL agents state space, the attacks presented target the RL agents action space which corresponds to actuators in CPS. Based on the state-action dynamics, the MAS is formulated as an optimization problem with decoupled constraints on the attack budget while the LAS operates by spreading the attack across the temporal and action dimensions.\par  
Just like in DNN's, black box attacks in RL are also more challenging than white box attacks because of the lack of information about the internal architecture and parameters of the target model. Although most research works have focused on the latter, there are recently few research that have shown results for black box attacks in RL \cite{zhao2020blackbox, gleave2019adversarial}. Inkawhich et al \cite{inkawhich2019snooping} posited that the state-of-the-art in black box attacks in RL which involves training a proxy agent and assuming that the adversary has a full knowledge of the environment is unrealistic. This is because, unlike in supervised learning where samples are from a static dataset, data generation in RL stems from the continuous exchange of the state, action and reward signals between the agent and state of the environment. They therefore proposed the snooping threat model, which operates with an assumption that the adversary, without access to the environment of the target resolves to eavesdropping on subsets of the RL signals at each time step. They therefore prove that by training proxy models on tasks similar to that of the target agent, adversarial samples can be crafted to compromise the performance of an RL agent. Interestingly, Pattanaik et al \cite{pattanaik2017robust} proposes adversarial attacks for RL and then leverage on the proposed attacks to improves the robustness of DRL to parameter uncertainties.
\par
Recently, other attacks on RL algorithms have been proposed. Our focus have been on methods that have received relatively wider recognition by the research community. For easy reference, Table III presents a summary of the attacks on RL. In the next section, we will look at techniques for developing robust and resilient RL models.

\subsection{Summary and lessons learnt}
\textcolor{black}{We have seen that recent trends in ML security have focused principally on resiliency of ML models to black-box adversarial attacks and the transferability of these attacks to other supervised learning models. However, the effect of adversarial attacks on unsupervised ML algorithms have not been well investigated. Some of the factors responsible for this neglect is the difficulty in defining what constitutes an adversarial example for clustering algorithms as a result of the absence of labels. Moreover, the inherently ad-hoc nature of unsupervised ML algorithms (such as clustering algorithms) also contribute to the factors that make adversarial ML a relatively more difficult task in comparison to supervised learning algorithms. Chhabra et al. \cite{chhabra2019strong} in an attempt to address these concerns developed an iterative black-box adversarial attack whose goal is to craft adversarial examples that fooled four clustering algorithms. Moreover, they studied the issue of adversarial attacks transferability in unsupervised ML models. Furthermore, the same authors in \cite{2020suspicion} proposed a definition for adversarial examples in clustering algorithms. Consequently, they presented a powerful black-box adversarial attack algorithm against clustering algorithms for linearly separable clusters. Their simulation results showed that the proposed method succeeded in generating adversarially perturbed samples by changing the decision boundary and therefore ensuring that the examples were mis-clustered.}\par 
\textcolor{black}{With these recent discoveries, it is expected that research into adversarial attacks on unsupervised learning algorithms will be given more attention by researchers. However, whether this attention will be as intensive as the supervised learning algorithms remains a doubt as ML algorithms in themselves have become very popular for their role in classification. We therefore posit that attention will focus on supervised and reinforcement learning than on unsupervised learning algorithms, especially for CPS applications.}

\begin{table*}[]
	\setlength\extrarowheight{4pt}
	\centering
	\caption{Summary of Research on Adversarial Attacks on Reinforcement Learning }
	\label{}
	\begin{tabularx}{\textwidth}{|R|R|R|R|}
		\toprule
		\textbf{Research} & \textbf{Name of Attack} & \textbf{Method of crafting adversarial examples} & \textbf{Contribution}\\
		\hline
		Adversarial Attacks on Neural Network Policies \cite{huang2017adversarial}  & White Box / Black Box & Fast Gradient Sign Method & The FGSM in \cite{goodfellow2014explaining} is extended to the DRL domain to successfully fool agents. They also showed that the transferability property also holds in RL.   \\
		\hline
		
		Tactics of Adversarial Attacks on DRL agents\cite{lin2017tactics} & Strategically-timed \& Enchanting attacks & Carlini \& Wagner  & Limits the duration of an attack to achieve optimal effects thus guaranteeing stealthiness. Also capable of luring an agent into a targeted state. \\
		\hline

		Delving into Adversarial Attacks on Deep Policies \cite{kos2017delving}  & Value Function (VF) attack & Fast Gradient Sign Method & The frequency of injection of adversarial examples into the agent is significantly reduced by guiding the attacker to attack at crucial moments using computations based on VF \\
		\hline
		
		Spatiotemporally Constrained Action Space Attacks on DRL agents\cite{yeow2019spatiotemporally} & White-box Myopic Action Space (MAS)/White-box Look-Ahead Action Space (LAS) & Similar to FGSM, with standard gradients computed & The adversarial attack focus on the agent's action space (which represents actuators in CPS) as against other works that have focused more on the agent's state space\\
		\hline	
		
		Vulnerability of Deep Reinforcement Learning to policy induction attacks  \cite{behzadan2017vulnerability} & Policy induction attacks & FGSM \& JSMA & First to establish the possibility of adversarial examples and their transferability in Deep Q-Networks. Using a game scenario, an attack mechanism to exploit perturbation and transferability to attack the policy was proposed.   \\
		\hline
		
		Targeted Attacks on DRL Agents through adversarial observations\cite{hussenot2019targeted} & Per-observation attack \& Universal-masks attacks & Fast Gradient Sign Method & The attacks here are targeted and also aimed at observation of the environment and not the internal state of the agent as usually considered. The attacks are also constant and not per-observation as usual.\\
		\hline
		
		Snooping Attacks on Deep Reinforcement Learning\cite{inkawhich2019snooping} & Snooping threat models & Fast Gradient Sign Method & Unlike other works, the adversary has no personal interaction with the environment; it eavesdrops the action and reard signals exchanged and then launches an attack on an agent by training proxy models on related tasks and then transferring the attacks.  \\
		\hline
		
		Stealthy and Efficient Adversarial Attacks against Deep Reinforcement Learning\cite{sunstealthy} & Critical point attack and Antagonist attack & Carlini \& Wagner & The critical point attack can predict the environment states and locate the critical moments to incur the most damage while the antagonistic attack can automatically identify the optimal attack strategy using the least attack cost.  \\
		\hline
		Sequential Attacks on Agents for Long-Term Adversarial Goals\cite{tretschk2018sequential}  & Sequential attacks & Adversarial Transformer Network (ATN)& Sequential attacks are applied at test time with adversarial examples crafted using ATN with the ultimate aim of manoeuvring a target policy network to pursue an adversarial reward.  \\
		
		\bottomrule
	\end{tabularx}
\end{table*}

\section{Secure/Resilient ML}
Previous sections of this paper have shown that ML algorithms enhance resiliency in CPS. Specifically, the communication, control and computing tasks can be been made robust to ensure their continuous operation during adversarial attacks. However, a recent research problem is that ML algorithms themselves are susceptible to attacks by the adversaries. Such attacks include data integrity attacks during the training or testing stages using data poisoning or the use of adversarial examples to fool the ML model to make wrong classifications. The use of GAN's have also made it easier to fool ML models. It is therefore pertinent that ML algorithms themselves must be made resilient to such attacks in order to guarantee safety and security of the systems where they are deployed. Therefore, due to the importance of ML to fields of cybersecurity, CPS and IoT, efforts must continue to ensure that these ML models are resilient to such attacks. Research efforts to achieve resilient ML is the focus of this section.
\subsection{Defense against adversarial attacks in DL}
Papernot et al.\cite{papernot2016distillation} suggested that certain requirements must be met in the quest to design defense mechanisms against adversarial perturbations. These include the need to limit impact on the architecture of the network, maintain the classification accuracy and speed of the target neural network. Although, research in this area is still at its infant stage, the interest is expected to rise significantly in the new decade as stated in many forecasts and white papers on the future of cybersecurity and AI. Various works are presently ongoing to address this challenge and some solutions have been proposed. \par 
This section therefore discusses the techniques or strategies for guaranteeing resilient ML. The majority of research presented focus on supervised learning tasks such as computer vision and image classification tasks using DNN's. From Figure 12, it is evident that defense against adversarial attacks can generally be achieved in three major ways. The first is a modification of the input through techniques like adversarial training\cite{goodfellow2014explaining}, transformation of the input through processes like compression and reduction of the bit-depth\cite{guo2017countering}, randomization of the data\cite{xie2017mitigating} and a regularization of the input gradient\cite{ross2018improving}. Next, defense against against adversarial attacks can also be achieved through a modification of the network structure through techniques such as defensive distillation\cite{papernot2016distillation}, multimodel-based defense\cite{srisakaokul2018muldef}, the addition of a detector sub-network\cite{metzen2017detecting} and also the addition of a high-level representative guided denoiser\cite{liao2018defense}. Finally, the objective function can also be modified by adding a stability term\cite{zheng2016improving}, adding a regularization term\cite{yan2018deep} and the stochastic activation pruning\cite{dhillon2018stochastic}. \par
The most popular methods of adversarial defense are discussed further in this section.
\subsubsection{Adversarial Training}
Adversarial training is one of the strongest methods of making ML algorithms resilient to adversarial attacks. In this method, examples are generated during the training of the ML model to harden it. Initially introduced by Szegedy et al. \cite{szegedy2013intriguing}, it was not fully implemented because of the challenge of generation of adversarial examples at the time. However, with the development of the FGSM, a fast method for generating adversarial examples, more extensive work was implemented and reported by Goodfellow et al.\cite{goodfellow2014explaining}. \par 
The adversarial training procedure therefore has a goal of minimizing the worst case error when an adversary introduces perturbations into the input data. Moreover, it can be likened to a form of active learning where the model has the ability to request labels on new points. Research has also shown that they are able to achieve high success rates. However, in a bid to improve the method and address certain drawbacks, variants of adversarial learning have since been developed to improve the resilient of ML algorithms to attacks.\par
Tramer et al.\cite{tramer2017ensemble} proposed ensemble adversarial training for adversarial defense of ML models. This model differs from the method initially proposed by separating the generation of the examples from the trained model. The method therefore enhances the training data and consequently the robustness of the target model to black-box attacks. Na et al.\cite{na2017cascade}, with a focus on unknown iterative attacks also proposed the cascade adversarial training to enhance the robustness of DNN's. Building on the principle of ensemble adversarial training, the cascade adversarial training uses already defended network for generation of iterative FGSM images,  in addition to the one-step adversarial images crafted from the network being trained. 
\subsubsection{Mitigation against adversarial examples through randomization}
Another method proposed for securing ML models against adversarial attacks using adversarial examples is the used of randomization\cite{xie2017mitigating}. According to the authors, carrying out randomization at inference time through random resizing and padding is capable of making systems resilient against adversarial attacks. The advantages of this method included the ability to make the network much more robust to adversarial images (especially for iterative attacks including white-box and black box attacks), non-requirement for training or fine-tuning, requirement for few computations thus having no real effect on computational complexity, and their compatibility with different network structures and adversarial defense methods. 
\subsubsection{Defensive Distillation}
Defensive distillation is another principle for guaranteeing resiliency of ML algorithms used in CPS. First proposed by Papernot et al.\cite{papernot2016distillation} as a defense mechanism for adversarial attacks especially within the context of DNN's, the principle had earlier been suggested by \cite{ba2014deep} and formally presented by \cite{hinton2015distilling}. Primarily developed for solving computational complexity challenges in DNN's, it involves the transfer of knowledge from one model to another. There are three training steps in defensive distillation. First, a network is trained using the normal known techniques. Then, it is evaluated by every instance of the training set to produce soft labels. Lastly, a second network known as the distilled network is trained on the soft labels generated in the second step.
Papernot et al. further proved the effectiveness of defensive distillation in mitigating adversarial attacks created using the FGSM and the Jacobian-based iterative approach by providing experimental results\cite{papernot2016effectiveness}.\par
However, as newer adversarial examples were crafted, the challenges of defensive distillation were also unmasked. The major discovery is that it did not work well in detecting certain adversarial examples\cite{papernot2017practical, carlini2017towards}. The authors in \cite{carlini2017towards} posited that slightly modifying a standard attack such as FGSM, defensively distilled networks can still fall prey to adversarial examples. Specifically, it was observed that since distillation operates as a defense mechanism to adversarial examples by increasing the magnitude of the inputs to the softmax layer, a successful attack can be achieved by reducing the magnitude of the input.  Seeking to extend the principle of defensive distillation with the knowledge of its challenges as highlighted, the proponents of the technique suggested that it is more effective in white-box than in black-box attacks. They however concluded that the fact that the method does not require the defender to generate adversarial examples still makes it a method of choice. They also propose using defensive distillation together with adversarial training to improve ts efficiency.\cite{papernot2017extending}. \par
The use of defensive distillation as a technique for defense against adversarial attacks is still potent. However, there is a need for improvement so that new forms of attacks can be addressed. Furthermore, it was stated in \cite{carlini2016defensive, carlini2017towards} that although this method of adversarial defense can only be validated on available attacks, there is a need to understand the intrinsic qualities of the mechanism that determine its success and also investigate further by modifying other parameters with a goal to ascertain if the defensive duties will still be properly carried out. This serves as a proactive step to ensure that when parameters are modified, a defensive mechanism does not lose its power. 

\subsubsection{Gradient Masking}
Studies have shown that most of the methods used to construct adversarial examples depend on the knowledge of the gradient of the model to operate effectively. A method known as gradient masking was therefore proposed for protecting DNN's\cite{papernot2016towards}. Like the name implies, the gradient masking method is built around preventing the attacker from gaining knowledge of the gradient of the model. After gradient masking has been applied to a neural network, the result is usually a model that is smooth in specific neighborhoods of training points. Consequently, without a knowledge of the gradient of the model, the attacker will not know in what direction to perturb the input and thus succeed in making the model mis-classify. \textcolor{black}{Recently, newer methods that operate similar to gradient masking have been referred to as obfuscated gradients. The authors in \cite{athalye2018obfuscated} achieved this using gradient shattering, stochastic gradients and vanishing/exploding gradients. They therefore proposed three methods; backward pass differentiable approximation (BPDA), expectation over training (EOT) and re-parameterization to show that majority of the methods that depended on obfuscating gradients to prevent the generation of adversarial examples can be broken \cite{athalye2018synthesizing}.} \par
Furthermore, techniques earlier highlighted for launching successful attacks against black box models also affect the efficiency of gradient masking\cite{papernot2017practical}. Specifically, the ability of an adversary to train a substitute model which mimics the target model or model defended with gradient masking, and then use the input-output relationship of the substitute model to attack the targeted model weakens the argument for the use of gradient masking and other methods that only prevent gradient descent-based attacks as a defense mechanism in neural networks. 
\subsubsection{Detection Methods}
Adversarial attacks against ML models can also be detected. Methods for detecting such attacks begin by analysing the input sample to find out variations from the normal samples. Input samples which differ statistically from the normal examples are identified as adversarial. Differing from other methods like adversarial training and gradient masking that seek to harden DNN's by modifying the model, Xu et al.\cite{xu2017feature} proposed a new method known as feature squeezing for detecting adversarial examples. Observing that the feature input spaces are usually large and therefore an incentive for the construction of adversarial samples, they suggested that it is possible to squeeze or coalesce the unnecessary input features and so restrict the attack surface of the adversary. A comparison of the prediction of the model on samples before and after squeezing helps the model infer that a sample is adversarial if outputs are substantially different. Applying the proposed method in image classification, they use two feature squeezing methods to reduce the color bit depth of each picture and also for spatial smoothing. They concluded that the method is good because it is inexpensive, achieves high accuracy and few false positives and can also be used to complement other techniques. \par 
Other squeezing methods can also be used to protect DNN from adversarial attacks. 
Furthermore, Feinman et al.\cite{feinman2017detecting} proposed two features known as density estimates and Bayesian uncertainty estimates for detection of adversarial examples in DNN. Lu et al\cite{lu2017safetynet} also developed SafetyNet, a method for detecting and rejecting adversarial examples. The detector in this case ascertains that the image and depth map are consistent, hence identifying an adversarial example if a contrary situation occurs. Furthermore, Roth et al.\cite{roth2019odds} examined the change in features and log-odds when noise was applied to input samples to detect the operation of adversaries. They discovered that a characteristic direction is maintained during adversarial attacks while there is no specific direction in a normal scenario.  Other methods include the use of prediction difference\cite{guo2019detecting}, neural fingerprinting\cite{dathathri2018detecting}, stress response\cite{sun2019detecting}, feature distillation\cite{liu2019feature}, incremental learning of GAN's\cite{yi2019incremental}, convolutional filter statistics\cite{li2017adversarial} and Fisher information\cite{martin2019inspecting}.
\textcolor{black}{\subsubsection{Lessons learned}
Despite all the research work carried out in detecting adversarial examples and improving the resiliency of DNN to such attacks, there is still more work to be done. Carlini et al.\cite{carlini2017adversarial} posited that the detection of adversarial examples is not an easy task. A survey of ten of the proposed detection methods showed that they all can be fooled when new loss functions are constructed. The authors therefore conclude that the existing defense mechanisms have not been thoroughly subjected to tests and that research into the detection of adversarial examples is yet to reach a satisfactory point.} \textcolor{black}{Furthermore, in \cite{athalye2018obfuscated}, the same sentiments were reemphasized. The authors therefore called for more intensive threat models that do not limit activities of the attacker and also have an idea of the structure of the defense model.}     \par
\textcolor{black}{Few years after making this discovery, a study of literature still shows that there is still no defense method that can adequately withstand all attack mechanisms. Researchers who propose defense methods should therefore ensure to test them with as many attacks as possible and also carry out research into finding loopholes attackers can use to launch attack. However, we acknowledge that attack and defense mechanisms will continue in a game-like manner. Discovery of new defense methods will birth the launch of new attacks. The general consensus that attacks are relatively easier to construct than defense mechanisms calls for intensified efforts from researchers. We posit that researchers should focus on re-evaluating state-of-the-art defense methods before attempting to propose newer schemes. This will greatly enhance research in developing resilient ML. Furthermore, the potentials of adversarial training are great. However, issues of computational complexity and the time of reaction to attacks in real-life CPS applications need  for attention if the proposed defense methods will be deployed in such scenarios.}

\begin{figure*}
\includegraphics[width=\linewidth]{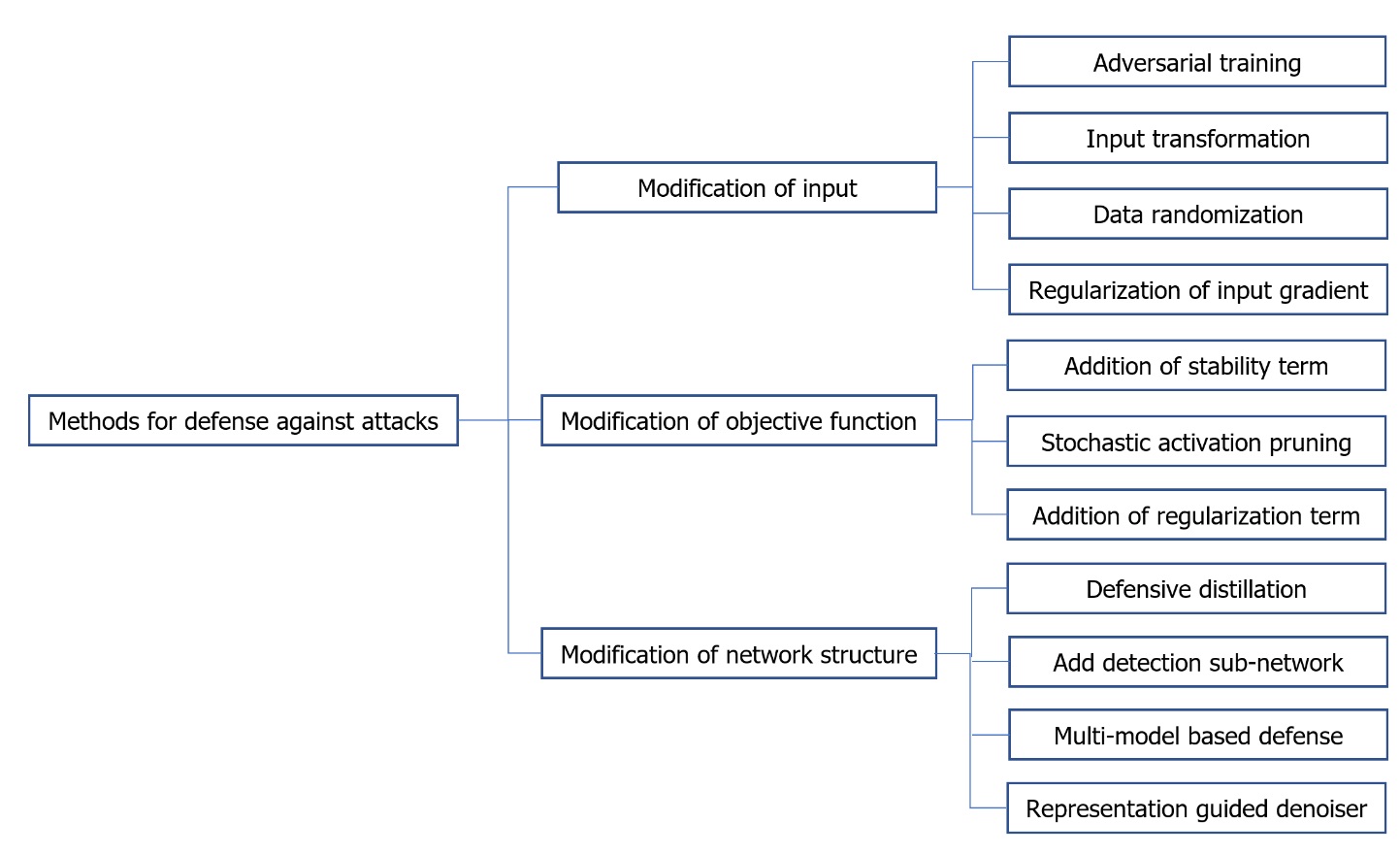}
\centering
\caption{Categorization of methods for defense against adversarial attacks}
\label{Figure 9}
\end{figure*}

\subsection{Defense against adversarial attacks in RL}
Having already discussed the possibility of attacking RL, the security of these systems is of concern to the researchers all over the world. The reality that adversarial attacks cannot be entirely isolated using access control methods is a call to develop algorithms that can operate despite these attacks. \textcolor{black}{Qu et al. in \cite{qu2020minimalistic} investigated the minimal requirements necessary to launch an adversarial attack in DRL. They considered cases where the attacker is only knowledgeable of the state and action, perturbs only a small number of pixels and attacks only some significant frames of the RL model. The simulations results show that attacking a single pixel which corresponds to 0.01\% of the state or attacking around 1\% frames totally fooled the trained policy using DQN. This research therefore shows that without developing robust defense against attacks, DRL cannot be used in critical infrastructure like in robotics and ITS.} \par Research on defense of DRL against adversarial attacks is still in its infant stage.  Many of the techniques for the resiliency of DRL stems from those developed for DNN's. Ilahi et al. \cite{ilahi2020challenges} classified defenses for adversarial attacks on DRL into adversarial training, robust learning, adversarial detection and defensive distillation. However, some of the researchers, in presenting their attacks recognized the challenge of defending systems against these attacks and therefore gave recommendations. The major method recommended for resiliency of DRL models is the adversarial training. With regards to images classification tasks using DNN, it has been proven that defensive distillation is not a reliable method for protecting ML models from attacks and thus it has not been considered for DRL by most researchers. The different variants of adversarial training suggested are presented first and then other methods are discussed. \par
Behzadan et al.\cite{behzadan2017whatever} investigated the resilience of DRL to training and test time attacks and concluded that the policies that are learnt with adversarial training can withstand test time attacks better. Havens et al.\cite{havens2018online} stated that most strategies for defending RL policies against adversarial attacks using adversarial training are usually in an off-line method. The disadvantage of this is that such strategies cannot adapt when the attack is online. Also, another demerit of other techniques is that the defense mechanisms are only effective for specific attacks. With these concerns in mind, the authors proposed the meta-learned advantage hierarchy (MLAH), an algorithm that detects and mitigates attacks on the state of the algorithm. The algorithm operates by using the advantage map, a metric estimated by a comparison of the expected return of a state to the observed return of some action to determine the presence of an adversary. The master agent, using the advantage map is able to switch between the nominal and adversarial sub-policies in correspondence to the attack scenario. The advantages of the MLAH algorithm include its online nature, ability to operate in the decision space and its effectiveness irrespective of the type of attack. However, the nature of the algorithm makes it computational intensive and increases the delay in detection because the target agent has to be fooled before the master agent can begin its defense procedure. Furthermore, the adversarially robust policy learning (ARPL) was proposed in \cite{mandlekar2017adversarially}. This algorithm, targeted at the defense of autonomous agents in physical domains like self-driving cars and robots uses adversarial agents during the training of RL agents to make them resilient to adversarial attacks in the form of changes in the environment. The authors start by proposing a method to generate adversarial perturbations that are plausible in physical systems. Thereafter, they proceed to use the ARPL to actively select the perturbations that are used to train the policy to make it more robust. Analysis of the effect of perturbations on performance in the presence of dynamics noise, process noise and observation noise makes this research well relevant to the resilient RL research. With future work such as the development of a theoretical justification for the algorithm and testing it on a physical robots, there is still enough room for research to make the algorithm deployed in real systems.\par 
For adversarial detection, Lin et al.\cite{lin2017detecting} proposed a defense method that both detects adversarial attacks in DRL and also provides suggestions on actions to be taken under such attacks. In this method, a visual foresight module is trained to detect the presence of adversarial examples by comparing the current action of the policy with the action generated by the same policy using a predicted frame. The Stochastic Activation Pruning (SAP) method proposed by Dhillon et al.\cite{dhillon2018stochastic} draws its inspiration from game theory. The SAP method prunes a random subset of activations and then scale up the ones left to makeup for the loss.\par

\textcolor{black}{Due to the challenges associated with using adversarial examples for improving the robustness of DRL models to adversarial attacks, researchers have now focused on methods that seek to certify or verify the robustness bounds. These methods depend on the approach proposed by Weng et al. and initially used for classification tasks \cite{weng2018towards} . However, Oikarinen et al. \cite{oikarinen2020robust } posited that the reason there are fewer research works that use this approach is the presence of challenges like an absence of a stationary training set and distinct right action for each state. Everett et al. \cite{everett2020certified} also towing the same line of certified adversarial robustness developed an online method for robustness of DRL algorithms. The proposed method; CARRL selected the action with the highest guaranteed lower bound Q-value during the execution process. The policy learnt also had an advantage of providing a certificate even when the certifier is unaware of salient details. The approach was tested with a DQN policy and confirmed to improve robustness when applied to pedestrian collision avoidance scenarios and a classic control task. Similarly, Wang et al. \cite{wang2019verification} used the same approach but  extended the research on robustness certification to a dynamic setting which is similar to the CPS. They therefore developed an algorithm for the certification of robustness of a DRL in a feedback control loop experiencing persistent adversarial attacks. Their method was confirmed to perform better than the conventional Lipschitz-based robust control approach, especially where the knowledge of the  model dynamics is unknown.} Other methods in this category were proposed in \cite{raghunathan2018certified, wong2018provable}\par 

\textcolor{black}{Oikarinen et al. \cite{oikarinen2020robust} however submitted that the authors in \cite{everett2020certified, wang2019verification} did not propose techniques for training more robust models. They therefore proposed the RADIAL-RL, a method that improves the robustness of DRL agents through the design of adversarial loss functions coupled with robustness verification bounds during the training process. They also proposed the greedy worst-case reward (GWC) which provides as a good estimate of the reward under the worst case sequence of adversarial attacks. By accounting for the importance of each action, GWC has an advantage over other approaches that only evaluate whether each single action is affected by input perturbations. 
This approach is similar to those proposed in \cite{fischer2019online, zhang2020robust}. Fischer et al. \cite{fischer2019online} proposed Robust Student-DQN (RS-DQN),  a method that uses both adversarial training and certified defense for the defense of DRL. The major contributions include splitting the DQN architecture into a student and a Q network. While the student network is used for exploration, the Q network is used for conventional training. The method therefore depends on imitation learning for a robust prediction of actions. However, the authors in \cite {zhang2020robust}, while acknowledging the efficiency of RS-DQN stated that it does not detail its behavior in environments without continuous action spaces. They therefore proposed the state-adversarial Markov decision process (SA-MDP), a method that does not depend on imitation learning and performs better in eleven test environments. The results showed that their method improved the adversarial robustness in PPO, DDPG and DQN agents.} \par

\textcolor{black}{Qu et al. \cite{qu2020defending} faulted the state-of-the-art approach of boosting adversarial robustness in DRL through policy distillation using adversarial training. They posited that by adding adversarial examples while the student policy is being trained, the robustness of the model suffers when it encounters a new attack. Moreover, this approach increases the computational cost of the model. To address this challenge, the authors proposed a new policy distillation method that does not include generation of adversarial examples during training for the defense of RL models against adversarial attacks. Specifically, they designed a policy distillation loss function that consists of the prescription gap maximization (PGM) loss and Jacobian regularization (JR) loss. The PGM loss maximizes both the probability of the action selected by the teacher policy and the entropy of the unwanted actions while the JR minimizes the norm of the Jacobian with respect to the input state. Theoretical and experimental analysis shows the accuracy of the new policy distillation and an improvement in robustness to attacks.}

\textcolor{black}{\subsubsection{Summary and lessons learned}
	Research into adversarial attacks and defense in RL is expected to continue to evolve in the coming years. However, from the research trend presented in this section, these developments will be hinged on some research findings. The first is the possibility of intermittent attacks that occur over a subset of time steps thereby making it further possible for an adversary to operate stealthily and efficiently. Second, the ability to lure an agent into taking actions that direct it towards an adversarial reward will continue to be a threat to the application of DRL models. Also, the development of more efficient methods of crafting adversarial examples like the ATN and the development of adversarial black-box attacks will further extend the frontier of research in DRL. All of these factors form critical concerns for the application of DRL in safety-critical systems like drones, self-driving cars and other CPS. 
	Compared to DNN's, adversarial defense for RL is still in its infant stage. However, as the applications of DRL in CPS and other systems continue to grow, the issue of defense will also attract a lot of attention. }\textcolor{black}{We agree with the authors in \cite{qu2020defending} that adversarial training for RL, aside from challenges of increased computational cost will suffer when new attacks are encountered and posit that similar to their research, novel defense methods that do not involve the generation of examples during training will be more efficient and practicable for uncertain environments.}\par \textcolor{black}{Finally, the issue of designing systems without a broad consideration about security and adversarial attacks right from the onset has posed as a setback to addressing the security concerns afterwards. Therefore, it is pertinent that security concerns must be factored into the design of systems to enhance their resiliency to adversarial attacks.}

\section{Open Research Challenges and Future Directions}
The fields of cybersecurity, CPS and AI will continue to evolve. The challenges experienced in applying these technologies to solve real life problems are of continuous research interest. A lot of investment is being made by the industry, government and other consortia to ensure that these problems are solved. Since research has become inter-disciplinary as CPS is in itself an inter-disciplinary field, there are some emerging technologies or developments that will contribute to actualizing AI-driven cybersecurity especially in CPS. Some of these are at very early stages but findings from white papers, blogs, and recent publications show that with deeper research interest into these topics of interest, they will without doubt influence the field positively. Furthermore, questions and issues of concern that should be brought to the fore as we design these systems are being discussed in this section. 

\subsection{Adaptive Adversarial Training for resiliency in ML}
Adversarial training was earlier discussed as one of the major methods of enhancing resiliency of ML models. However, although the perturbation of all inputs during adversarial training has an advantage of robustness and resiliency, it also has the disadvantages of cost as a result of computational complexity and its potential of leading to poorer generalizations. The principle of adaptive adversarial training where the input to be selected for perturbation are carefully selected in an adaptive manner has become a solution to this challenge. In \cite{balaji2019instance}, an instance adaptive adversarial training technique that carries out sample-specific perturbation margins around every training sample was proposed. Their results showed that with a marginal drop in robustness, generalization was achieved when the model was tested with unperturbed samples. 

\subsection{Context awareness for resiliency in ML}
Context awareness is the ability of a system or system component to gather information about its environment at any given time and adapt behaviors accordingly. Context awareness is also a technological driver for M2M (machine to machine) and IoT, ubiquitous computing and event-driven computing environments. An online context-aware ML algorithm for 5G millimeter-wave vehicular communications was proposed in \cite{8472783}. The algorithm sourced for sparse user location information, aggregated the received data and was thus able to learn and adapt to the environment. Resiliency of ML in CPS applications can therefore be satisfactorily achieved if the systems are context aware. 

\subsection{Federated Learning (FL) in Cybersecurity}
Recently, as a result of the big data revolution, data sets have become very large and the models have also become more complex thus making the training process more difficult. Federated Learning was proposed as the solution to this challenge. In edge devices such as mobile phones and IoT devices, there is a constraint in resources such as memory and processor and electric power. FL therefore gives them the capability to learn a shared model for prediction, while keeping the training data local. The advantages of this principle of decentralizing training models include privacy, security, regulatory and economic benefits \cite{DBLP:journals/corr/abs-1806-00582}. Doku et. al in \cite{doku2019towards} applied the principle of FL to determine the data relevance in big data. \newline In the nearest future, the principle of federated learning is therefore expected to be used to secure systems using ML.
\subsection{Distinguishing Malicious Attacks from Faulty Systems}
CPS systems comprise of a lot of components and sub-systems that also have the potential to fail. It is possible for some sensors to become faulty and produces wrong readings that result in a wrong output. However, because security systems that have been trained to identify the right behavior and the input of the system, and also be on the lookout for adversarial examples or data poisoning attacks and other signs of anomalous behavior and react, such system in seeking to achieve resiliency of the system will classify such a hardware fault as an attack and take appropriate moves. This would increase the computational power consumption of the system. The problem of distinguishing malicious attacks from failures of sensors and other devices in the CPS must therefore be solved even as we continue to seek to build systems that are both resilient using ML and have resilient ML algorithms.

\section{Conclusion}
In this paper, we have seen that the recent internet and telecommunication penetration has propelled technologies such as IoT and CPS. However, the growing interconnection of devices and things has widened the cyber surface and therefore led to a lot of cybersecurity concerns. With the increase in the success of cybersecurity attacks, the effects of such attacks financially and the damage they can have on infrastructure, new methodologies to complement the traditional methods of preventing such attacks must be explored. The potentials and challenges in applying AI and ML in cybersecurity have been thoroughly examined and the concerns and future directions have also been identified. Without a doubt, ML and AI will play a long role in securing our cyber space from attackers but there are challenges that need to be examined to ensure the overall success. The quest to overcome these challenges will make this interactions between ML and cybersecurity continuous with the ultimate goal of ensuring that ML models serve dominantly as a defense strategy and not an attack strategy.


\end{document}